\documentclass[manuscript]{emulateapj} 
\usepackage{hyperref}

\shorttitle{Magnetically Controlled Weather} 
\shortauthors{Batygin, Stanley \& Stevenson}

\begin{document}
 
\title{Magnetically Controlled Circulation on Hot Extrasolar Planets}  
\author{Konstantin Batygin$^{1}$, Sabine Stanley$^2$ \& David J. Stevenson$^3$} 
\affil{$^1$Institute for Theory and Computation, Harvard-Smithsonian Center for Astrophysics, 60 Garden St., Cambridge, MA 02138} 
\affil{$^2$Department of Physics, University of Toronto, 60 St. George St., Toronto, ON} 
\affil{$^3$Division of Geological and Planetary Sciences, California Institute of Technology, 1200 E. California Blvd., Pasadena, CA 91125}

\email{kbatygin@cfa.harvard.edu}

\begin{abstract}
Through the process of thermal ionization, intense stellar irradiation renders Hot Jupiter atmospheres electrically conductive. Simultaneously, lateral variability in the irradiation drives the global circulation with peak wind speeds of order $\sim$ km/s. In turn, the interactions between the atmospheric flows and the background magnetic field give rise to Lorentz forces that can act to perturb the flow away from its purely hydrodynamical counterpart. Using analytical theory and numerical simulations, here we show that significant deviations away from axisymmetric circulation are unstable in presence of a non-negligible axisymmetric magnetic field. Specifically, our results suggest that dayside-to-nightside flows, often obtained within the context of three-dimensional circulation models, only exist on objects with anomalously low magnetic fields, while the majority of highly irradiated exoplanetary atmospheres are entirely dominated by zonal jets.
\end{abstract}

\section{Introduction}
The last decade's discovery and rapid accumulation of the transiting extrasolar planetary aggregate has uncovered a multitude of previously unexplored regimes of various physical phenomena. Perhaps the first unexpected discovery was the existence of Hot Jupiters (i.e. gaseous giant planets that reside within $\sim 0.1$AU of their host star), which arose from the earliest exoplanetary detections \citep{1995Natur.378..355M, 1996ApJ...464L.147M}.  Accordingly, among the most intriguing novel theoretical subjects, is the study of atmospheric dynamics on highly irradiated planets. 

Today, it is well known that the orbital region occupied by Hot Jupiters can also be occupied by lower mass (including terrestrial) planets \citep{2012arXiv1202.5852B}. However, because of their higher likelihood of transit and comparative predisposition for characterization, Hot Jupiters remain at the forefront of the study of extrasolar atmospheric circulation \citep{2011exop.book..471S}. 

\subsection{Hydrodynamic Global Circulation Models}

Dynamic meteorology is a phenomenologically rich subject because of the lack of separation of physical scales. In other words, differences in microphysical nature of a given system can have profound effects on its macroscopic state. As a result, it comes as no surprise that circulation patterns on Hot Jupiters generally do not resemble those on Solar System gas giants \citep{2008ApJ...685.1324S,2009ApJ...700..887M}.

From a hydrodynamical point of view, the circulational modes of typical Hot Jupiter atmospheres differ in two principal ways, compared to Solar System gas giants. The first and most obvious difference is the energetics. Unlike the outer Solar System, Hot Jupiters reside in an environment where the incoming stellar irradiation completely dominates over the intrinsic planetary heat-flux. As a result, circulation patterns on Hot Jupiters are driven primarily by the congenital dayside-to-nightside temperature differences \citep{2011ApJ...738...71S}. Furthermore, concurrent with the cooling of the planetary interior, the top-down heating of the atmosphere ensures the onset of stable stratification in the observable ($P > 100$ bars) atmospheric region \citep{2002A&A...385..156G,2007ApJ...661..502B}. 

The second difference lies in the extent to which the atmospheres are rotationally dominated. While Solar System gas giants rotate rapidly (i.e. $\mathcal{T}_{\rm{Jup}} \simeq \mathcal{T}_{\rm{Sat}} \simeq 10$ hours), Hot Jupiters are thought to rotate pseudo-synchronously with their orbital periods (i.e. $\mathcal{T}_{\rm{HJ}} \sim 3 - 5$ days) as a result of tidal de-spinning (see \citet{1981A&A....99..126H}). This implies that although still dynamically significant, rotation alone does not exhibit commanding control over the atmospheric flow. 

Since the discovery of the first transiting extrasolar gas giant HD209458b \citep{2000ApJ...529L..45C, 2000ApJ...529L..41H}, numerous authors have explored the atmospheric dynamics of Hot Jupiters with a variety of numerical techniques. Because of the inherent differences in the frameworks of the simulations, today, there exists a hierarchical collection of results that correspond to variable degrees of sophistication. On the simpler end of the spectrum are 2D shallow-water simulations \citep{2008ApJ...675..817C,2003ApJ...587L.117C, 2008ApJ...674.1106L, 2007ApJ...657L.113L} while the more intricate global circulation models (GCM's) include solvers of the 3D ``primitive" equations \citep{2005ApJ...629L..45C,2008ApJ...685.1324S,2009ApJ...700..887M, 2011MNRAS.418.2669H} as well as the 3D fully compressible Navier-Stokes equations \citep{2008ApJ...673..513D,2012arXiv1211.1709D}. Simultaneously, various groups have gone to different lengths in their treatment of radiative transfer, with exploited models ranging from simple prescriptions such as Newtonian cooling \citep{2008ApJ...685.1324S} to double-gray \citep{2012ApJ...750...96R} and non-gray \citep{2009ApJ...699..564S} schemes. An important step towards delineating the correspondence among results obtained with different solvers has been recently performed by \citet{2011MNRAS.tmp..370H}.

Although there are quantitative differences in the results generated by different GCM's, there is general agreement on the qualitative features of the circulation. Specifically, there are three aspects of interest. First, super-rotating zonal jets exist in all simulations. Their number (and naturally, the widths) ranges between 1 and 4, depending on the model (see \citet{2009ApJ...699..564S}), but the relative sparsity of the jets compared to Jupiter and Saturn is understood to be a result of diminished rotation rate \citep{2002A&A...385..166S}. Moreover, in a recent study, \citet{2011ApJ...738...71S} showed that the formation of jets is ordained by the interaction of the atmospheric flow with standing Rossby waves that in turn result from the strong difference in the radiative forcing between the planetary dayside and the nightside. 

Second, the characteristic wind speeds produced by different models are consistent within a factor of a few, and are generally in the $\sim$ km/s range. This is likely a direct result of the overall similarity in the force-balance setup within the models. Specifically, \citet{2011exop.book..471S} argue that near the equator, the horizontal pressure-gradient acceleration caused by the asymmetric irradiation is balanced by the advective acceleration. Meanwhile, Coriolis force takes the place of advective acceleration as the primary balancing term in the mid-latitudes. Both force-balances yield $\sim$ km/s as the characteristic wind speeds, in agreement with the numerical models. 

\begin{figure}
\includegraphics[width=1\columnwidth]{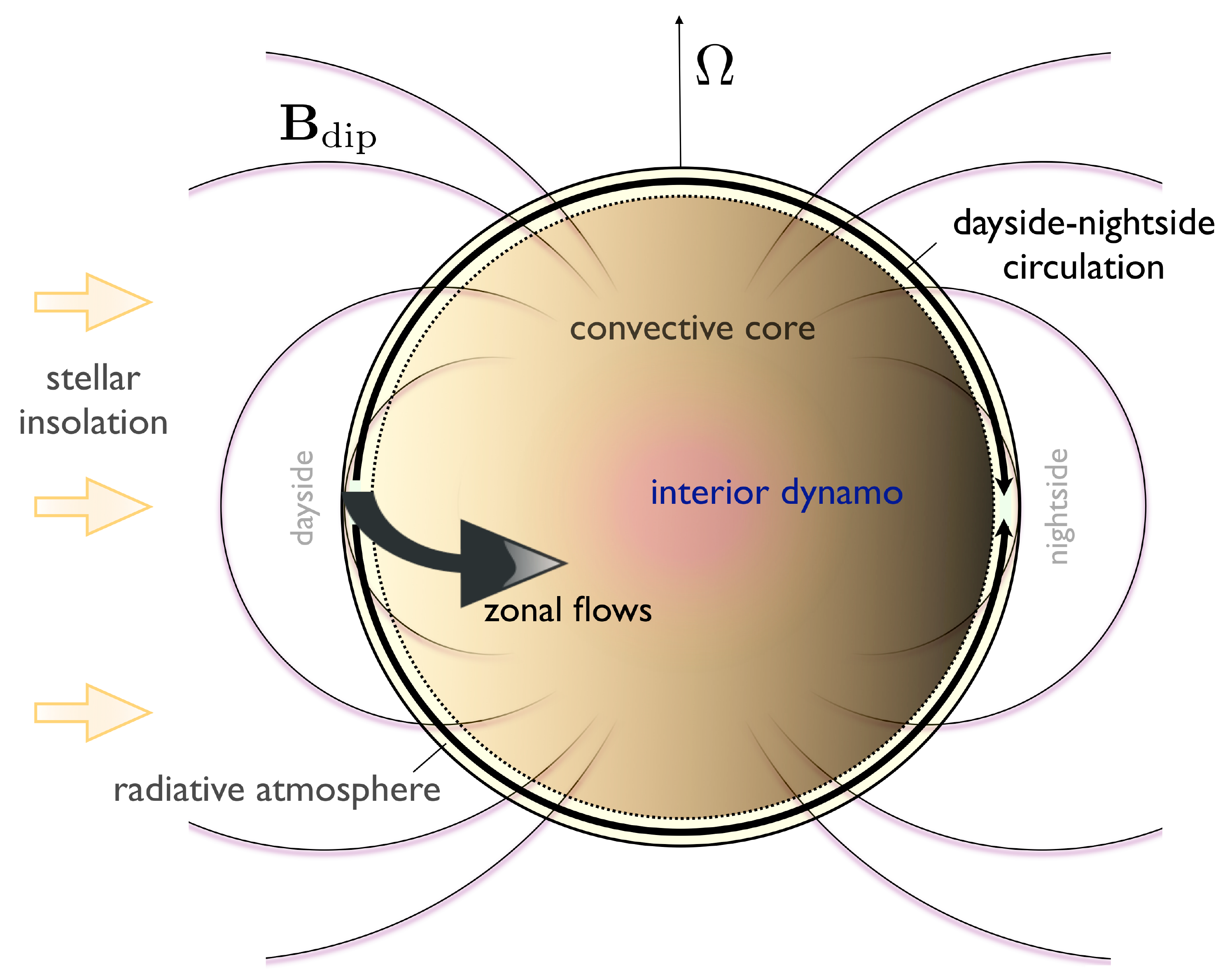}
\caption{A schematic diagram of the problem considered in this work. A spin-pole aligned dipole magnetic field is thought to arise from a dynamo operating in the deep interior of the planet. As a result of thermal ionization of Alkali metals present in the radiative atmosphere, the interactions between high-velocity flows and the background field lead to non-trivial corrections to the hydrodynamic solution of the global atmospheric circulation. It is likely that the topologically more complex dayside-to-nightside flows in the upper atmosphere ($P \sim $ mbar) are more affected by magnetohydrodynamic effects than the zonal flows that reside in the deep atmosphere.}
\label{prettywoman}
\end{figure}

Finally, in GCMs that resolve the vertical structure of the atmosphere (e.g. \citet{2008ApJ...685.1324S,2009ApJ...700..887M, 2011MNRAS.418.2669H}) eastward jets consistently dominate the lower atmosphere while the upper atmosphere is characterized by more or less symmetric dayside-to-nightside circulation. In other words, winds originate at the sub-solar point and flow to the anti-solar point over the terminator in the upper atmosphere. The transition between the circulation patterns takes place at $P \sim 0.1 - 0.01$ bars and is a consequence of the substantial reduction of the radiative time constant with diminishing pressure \citep{2005A&A...436..719I}. In particular, because the radiative adjustment timescale is much shorter than the advective timescale in the upper atmosphere, the flow is unable to perturb the temperature structure away from radiative equilibrium significantly. Figure (\ref{prettywoman}) depicts a schematic representation of the characteristic features of atmospheric circulation on Hot Jupiters. 

\subsection{Magnetically Dragged Global Circulation Models}

\begin{figure}
\includegraphics[width=1\columnwidth]{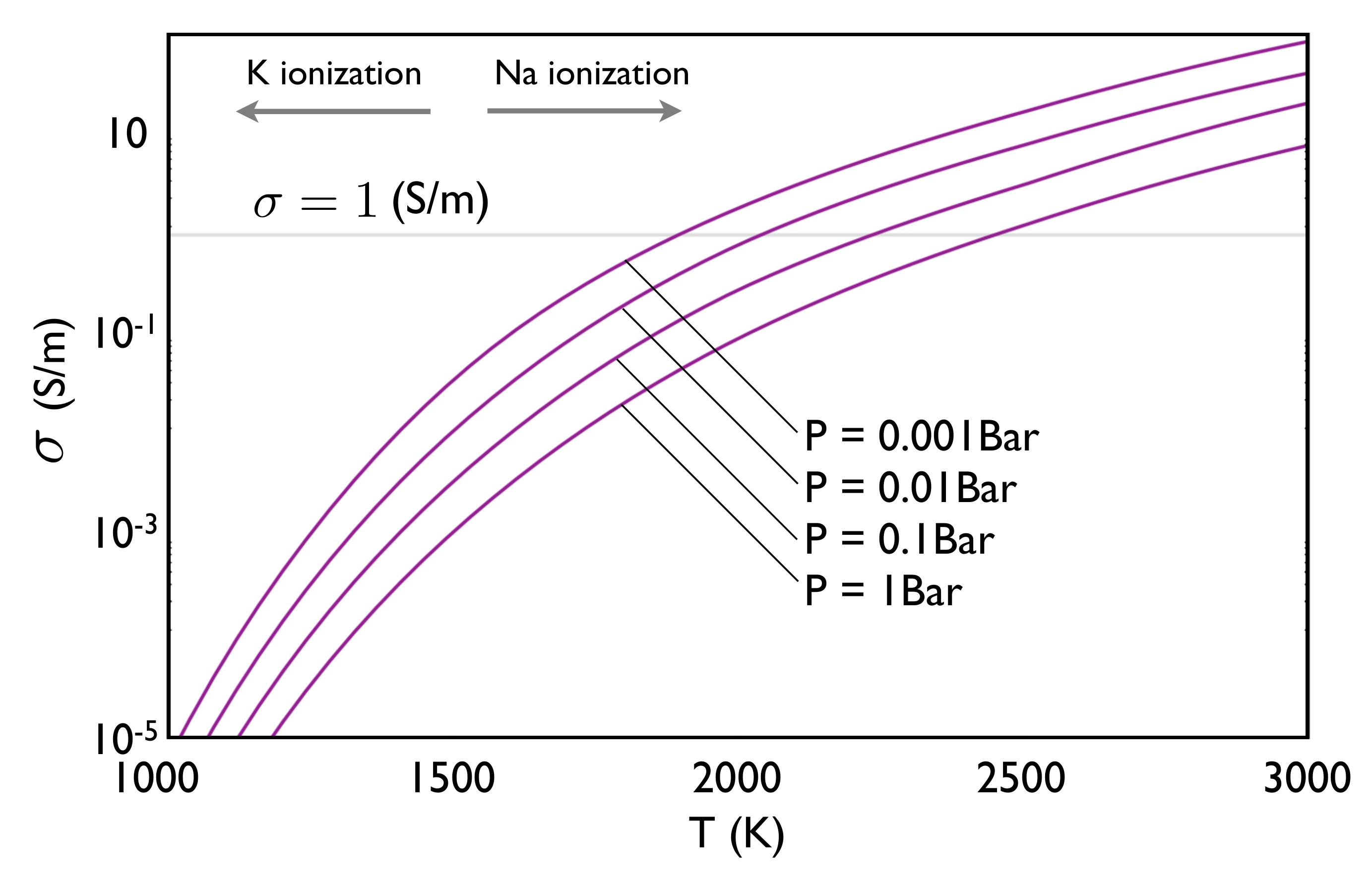}
\caption{Electrical conductivity at various pressure levels in a typical Hot Jupiter atmosphere. The conductivity arises as a result of thermal ionization of Alkali metals and the curves are computed as done in \citep{2010ApJ...714L.238B}. While the ionization of K dominates at lower temperatures, it saturates at $T \sim 1500$K, giving way to Na as the primary additional source of free electrons. Note that the conductivity is only weakly dependent on density.}
\label{sigma}
\end{figure}

There exists another important, distinctive feature of Hot Jupiter atmospheres, namely their non-negligible electrical conductivity (see Figure \ref{sigma}). Electrical conductivity in Hot Jupiter atmospheres does not originate from the ionization of H or He, but rather from the stripping of the valence electrons belonging to alkali metals such as K and Na \citep{2010ApJ...714L.238B, 2010ApJ...719.1421P}. While these elements are thought to be present in trace abundances (e.g. [K]/[H] $\sim 10^{-6.5}$, [Na]/[H] $\sim 10^{-5.5}$) \citep{1999ApJ...519..793L}, temperatures of $\sim 2000$K at upper atmospheric pressures, lead to total and partial ionization of K and Na respectively. In fact, at mbar levels, the conductivity can reach values as high as $\sigma \sim  1$ S/m \citep{2011ApJ...738....1B, 2012ApJ...750...96R,2012ApJ...748L..17H}. Furthermore, it is generally expected that much like solar system gas giants, Hot Jupiters posses interior dynamos, that produce surface fields comparable to, or in slight excess of Jupiter's field\footnote{Although, it is possible that bodies with diminished internal heat fluxes (see e.g. \citet{2007ApJ...668L.171B}) may have comparatively lower fields. Unfortunately at present, strengths and morphologies of exoplanetary magnetic fields remain observationally elusive \citep{2013ApJ...762...34H}.} (e.g. $B \sim 3 - 30$ Gauss) \citep{2003E&PSL.208....1S,2009Natur.457..167C}. Consequently, there is a distinct possibility that atmospheric circulation on Hot Jupiters may be in part magnetically controlled. That is to say, highly irradiated atmospheres may be sufficiently conductive for the Lorentz force to play an appreciable, if not dominant role in the force-balance. 

Realizing the potential importance of the coupling between the mean flow and the planetary magnetic field, \citet{2010ApJ...719.1421P} modeled the Lorentz force as a Rayleigh drag (a velocity-dependent force that opposes the flow) and incorporated it into the GCM previously utilized by \citet{2010ApJ...713.1174M}. This effort was later amended by \citet{2013ApJ...764..103R}, who also modeled the Lorentz force as a Rayleigh drag but self-consistently accounted for spatial variability in the electrical conductivity (by extension the drag timescale) in the weather layer. The results obtained with dragged GCMs exhibit significant differences in the obtained flow velocities relative to the standard GCMs. Namely, \citet{2010ApJ...719.1421P} found a factor of $\sim 3$ decrease in the peak wind speeds as the background dipole magnetic field was increased from $B_{\rm{dip}} = 3$ G to $B_{\rm{dip}} = 30$ G, while \citet{2012ApJ...750...96R} found a similar decline in the jet speeds as the field was increased from $B_{\rm{dip}} = 0$ G to $B_{\rm{dip}} = 10$ G. The magnetic limitation of the peak wind speeds is of considerable importance as it may prevent the global circulation from approaching a super-sonic state (note that the characteristic sound speed is order $c_s \sim \sqrt{k_{\rm{B}} T / \mu} \sim 3$ km/s, where $k_{\rm{B}}$ is Boltzmann's constant, $T$ is the temperature, and $\mu$ is the mean molecular weight), thereby inhibiting the formation of shocks.

\subsection{The Necessity for Magnetohydrodynamic Circulation Models}

Although dragged 3D GCMs clearly highlight the quantitative importance of the magnetic effects in Hot Jupiter atmospheres, they fail to accentuate significant qualitative differences in the obtained flows. Specifically, much like conventional GCMs, magnetically dragged GCMs still show deep-seated zonal jets, overlaid by complex flow patterns that intersect the poles of the planets. This lack of qualitative differences may arise from two distinct possibilities. The first is that beyond diminishing the peak wind speeds, the background magnetic field has little effect on the global circulation. In actuality, this may very well be true for pressure levels where the circulation is dominated by zonal jets, because of the geometrical simplicity of the flow-field interactions. Indeed, the coupling between the zonal flow and the pole-aligned background dipole field is azimuthally symmetric: differentially rotating jets convert the poloidal field into toroidal field \citep{2008Icar..196..653L, 2010ApJ...714L.238B}. As will be discussed in detail below, beyond the reduction of velocities, this conversion poses few dynamical ramifications for the jets. Furthermore, owing to higher pressure and somewhat diminished temperatures compared with the upper atmosphere (and the associated decrease in conductivity), the zonal jets may reside in the kinematic regime, where the effects of the Lorentz force are modest \citep{2011ApJ...738....1B, 2012ApJ...754L...9M}.

The second possibility is that although in reality the interactions with the background field impel the circulation to strongly deviate from its purely hydrodynamical counterpart, the procedure of modeling the Lorentz force as a Rayleigh drag does not capture the essential features of the dynamics. This is likely true in the upper atmosphere, where azimuthal symmetry is broken and the flow takes on a more topologically complex form. After all, in such a setting there is no requirement for the Lorentz force to simply oppose the flow everywhere, as is done by Rayleigh drag. Thus, there is a distinct possibility that previous modeling efforts have consistently misrepresented the circulation patterns of the upper atmospheres of Hot Jupiters. Accordingly, the investigation of this possibility is the primary purpose of this work.  

A statistically sound comparison between theoretical models and observations requires the incremental decrease in the goodness of fit to outweigh the cost of introducing new degrees of freedom into the model (see \citet{2000SAOPP...2.....R} for an in-depth review). Within the context of extrasolar planets, the limitations in observational capabilities and the quality of the data render the construction of highly sophisticated models unjustified \citep{2012ApJ...749...93L}. Although a rigorous comparison with observational data is not the focus of this paper, our modeling efforts will lie in the same rudimentary spirit. In other words, here we shall focus on understanding the qualitative, rather than quantitative nature of the circulation. Numerous simplifying assumptions will be made and the representation of the flow (including flow velocities, dayside-to-nightside temperature differences, etc) should only be viewed as approximate. However unlike all previous works on the subject, the model we shall utilize will remain self-consistently magneto-hydrodynamic (MHD). In taking this approach, we hope to successfully capture the essential features of magnetic effects in highly irradiated planetary atmospheres.

The paper is organized as follows. In section 2, we describe the equations inherent to our numerical GCM and reproduce the main features of Hot Jupiter atmospheric flows in the purely hydrodynamic regime. In section 3, we discuss the qualitative features of the atmospheric flows, treating the Lorentz force as a hydrodynamic drag. Specifically, we develop an analytical theory for magetically-dragged circulation patterns in the upper atmosphere and test the resulting scaling law against numerical simulations with enhanced viscosity and explore the effects of varying the radiative timescale. In section 4, we introduce a pole-aligned background magnetic field and demonstrate the transition of the upper atmosphere's dayside-to-nightside circulation into a globally zonal state with the onset of the background field. We conclude and discuss our results in section 5. 

\section{Numerical Global Circulation Model}

The Hot Jupiter GCM we have adopted here is a  variant of the numerical geodynamo model constructed by \citet{1999JCoPh.153...51K}. Since its conception, the model's versatility has been exploited extensively to explain the geodynamo \citep{1999JCoPh.153...51K,2002GeoJI.151..377D}, the ancient Martian dyanamo \citep{2008Sci...321.1822S}, Mercury's thin-shell dyanamo \citep{2005E&PSL.234...27S, 2007SSRv..131..105Z}, Saturn's dynamo \citep{2010GeoRL..37.5201S}, as well as dynamos of Uranus \& Neptune \citep{2004Natur.428..151S,2006Icar..184..556S}.

The details of the implementation of the model and the utilized numerical methods are throughly described by \citet{1999JCoPh.153...51K}. Here, rather than exhaustively restating the particularities of the framework, we limit ourselves to presenting the set of equations under consideration and the underlying assumptions, while referring the interested reader to \citet{1999JCoPh.153...51K} for further information.

\subsection{The Governing Equations}

\paragraph{Momentum.} The circulation model solves the Navier-Stokes equation for an electrically conductive, Boussinesq fluid
\begin{equation}
\label{NavierStokes}
\frac{D \mathbf{v}}{D t} = - 2 \mathbf{\Omega} \times \mathbf{v}  - \frac{\nabla \mathcal{P}}{\bar{ \rho} } + \frac{\delta \rho}{\bar{\rho}} \mathbf{g} + \frac{\mathbf{J} \times \mathbf{B} }{\bar{\rho}} + \nu \nabla^2 \mathbf{v}
\end{equation}
in a rotating spherical shell of finite thickness. Here, $D/Dt = \partial/\partial t + \mathbf{v} \cdot \nabla$ is the material derivative, $\mathbf{v}$ is the velocity vector, $ \mathbf{ \Omega } = (2 \pi / \mathcal{T})  \mathbf{\hat{z}}$ is the rotation vector, $\mathcal{P}$ is the modified pressure, $\rho$ is the density, $\mathbf{J}$ is the current density and $\nu$ is the kinematic viscosity. The bar denotes an average, whereas $\delta$ denotes the perturbation away from the background state. The density and temperature are related to the pressure through the ideal gas equation of state\footnote{In practice, the equation of state only enters the Boussinesq equations indirectly, through the thermal expansion coefficient. }:
\begin{equation}
P = \frac{\rho}{\mu} k_{\rm{B}} T,
\end{equation}
where $P$ is the total pressure. The dynamic domain where the equation is solved is confined above a rigidly rotating spherical shell. We denote the inner and outer radii of the atmosphere as $r_1 < r_2$ respectively.

\paragraph{Continuity.} The model is formally 3D and the vertical component of the motion enters into the continuity equation:
\begin{equation}
\label{continuity}
\nabla \cdot \mathbf{v} = 0.
\end{equation}
However, the nearly-constant density, incompressible fluid approximation prevents us from self-consistently modeling a radially extensive atmosphere. Indeed, the atmospheric density does not change, except by thermal expansion/contraction\footnote{Recall that for an ideal gas, the coefficient of thermal expansion is $\alpha = (1/V)(\partial V/ \partial T)_P = 1/T$.}:
\begin{equation}
\frac{\delta \rho}{\bar{\rho} } =  - \frac{\delta T}{\bar{T}}.
\end{equation}
Consequently, we limit the thickness of the atmosphere to a single scale-height in our simulations: $r_2 - r_1 = \mathcal{H} = k_{\rm{B}} \bar{T} / \mu g$, where $g$ is the acceleration due to gravity. Additionally, we set $r_1 = \mathcal{R}$, the radius of the planet. 

Generally, because of the above-stated reasons, our model should be viewed as more closely related to the 2D shallow-water GCMs \citep{2003ApJ...587L.117C, 2008ApJ...674.1106L} rather than the family of 3D models \citep{2009ApJ...699..564S, 2010ApJ...719.1421P, 2012ApJ...750...96R}. However, proper treatment of the of the induction equation (see below) in absence of pre-described symmetry requires the model to remain 3 dimensional. 

\paragraph{Energy.} The energy equation, that governs the temperature, inherent to the model reads:
\begin{equation}
\frac{D T}{D t}  = \kappa \nabla^2 T,
\end{equation}
where $\kappa$ is the coefficient of thermal diffusivity (kept constant throughout the computational domain). Strictly speaking, this equation governs diffusive heat flux and (in direct interpretation) is unsuitable for modeling a medium where energy transport is accomplished mainly by radiation. This is because the above equation rests on the approximations of short photon mean free path and the neglect of the temperature-dependence of the opacity (in the context of the Boussinesq treatment employed here, the latter makes sense but the former breaks down at pressure levels corresponding to optical depth of order unity, allowing for only a crude approximation to reality). However, shall it be possible to relate $\kappa$ to radiative properties of the gas, the above energy equation can still be used to effectively mimic the appropriate heat transport. 

In a radiatively-dominated atmosphere, the correct energy equation reads \citep{1992phcl.book.....P}
\begin{equation}
\label{heatequation}
\frac{D T}{D t}  = \frac{1}{\bar{\rho} c_p} \nabla \cdot \mathcal{F},
\end{equation}
where $c_p$ is the specific heat capacity at constant pressure and $\mathcal{F}$ is the radiative heat flux. The expression for the radiative heat flux reads \citep{1968psen.book.....C}:
\begin{equation}
\mathcal{F} = \frac{4 \sigma_{\rm{sb}} \bar{T}^3}{3 \bar{\rho} \psi} \nabla T,
\end{equation}
where $\sigma_{\rm{sb}}$ is the Stefan-Boltzmann constant and $\psi$ is the opacity. At this point, the relationship between $\kappa$ and the atmospheric temperature, density, opacity, and heat capacity is obvious. However, before proceeding further, let us recall that to an order of magnitude, the Newtonian cooling timescale $\tau_{\rm{N}}$ is given by the ratio of the atmosphere's excess heat content to its excess radiative flux:
\begin{equation}
\tau_{\rm{N}} \simeq \frac{\bar{\rho} c_p \mathcal{H}}{4 \sigma_{\rm{sb}} \bar{T}^3}.
\end{equation}
Consequently, we may express:
\begin{equation}
\label{kappa}
\kappa = \frac{4 \sigma_{\rm{sb}} \bar{T}^3}{3 \bar{\rho}^2 c_P \psi} \simeq \frac{\mathcal{H}^2}{\tau_{\rm{N}}},
\end{equation}
where we have implicitly assumed an infrared optical depth of order unity at the pressure-level of interest. The relationship between $\kappa$ and $\tau_{\rm{N}}$ is convenient, as it can be related to previous works. In particular, \citet{2008ApJ...685.1324S} have calculated $\tau_{\rm{N}}$ using a state of the art radiative transfer model and tabulated the results on a pressure-temperature grid. Here, we utilize their computations as a guide in estimating the thermal diffusivity. 

Note that we could have arrived at the relationship (\ref{kappa}) more intuitively by dimensional analysis. Specifically, noting that the radial extent of the atmosphere is much smaller than the lateral extent, the relevant length scale is the vertical scale-height, $\mathcal{H}$. Meanwhile, because the heat transport is primarily radiative, $\tau_{\rm{N}}$ is clearly the relevant timescale. Bearing in mind the units of diffusivities (i.e. m$^2$/s), equation (\ref{kappa}) naturally emerges as an estimate. 

\paragraph{Magnetic Field.} The evolution of the magnetic field is governed by the induction equation \citep{1978mfge.book.....M}
\begin{equation}
\label{induction}
\frac{\partial{\mathbf{B}}}{\partial t} = \eta \nabla^2 \mathbf{B} + \nabla \times (\mathbf{v} \times \mathbf{B}),
\end{equation}
where $\eta = 1/\mu_0 \sigma $ is the magnetic diffusivity (kept constant throughout the computational domain) and $\mu_0$ is the permeability of free space. Meanwhile, the absence of magnetic monopoles implies a divergence-free magnetic field: 
\begin{equation}
\label{div0}
\nabla \cdot \mathbf{B} = 0.
\end{equation}
Once the structure of $\mathbf{B}$ is known, the current density (within the MHD approximation) is given by 
\begin{equation}
\label{J}
\mathbf{J} = \frac{1}{\mu_0} \nabla \times \mathbf{B}.
\end{equation}
At this point, the full set of governing differential equations is presented. Paired with a matching set of boundary and initial conditions, the system can be integrated forward in time self-consistently. 

The equations are solved using a mixed spectral--finite difference algorithm and following \citet{1999JCoPh.153...51K}, the spherical harmonic decomposition is taken up to degree $\ell_{\rm{max}} = 33$ in the latitude $m_{\rm{max}} = 21$ in the azimuthal angle. The computational domain is broken up into 64 radial shells. The model is integrated forward in time until equilibration in the thermal, kinetic and magnetic energies is attained.

\subsection{Boundary and Initial Conditions}

The physical parameters employed in the numerical experiments we report are loosely based on the planet HD209458b \citep{2000ApJ...529L..45C}. Aside from being the first extrasolar planet found to transit its host star, it has become a canonical example used in the studies of Hot Jupiter atmospheres \citep{2007ApJ...668L.171B, 2010Natur.465.1049S}. To this day, (along with HD189733b \citep{2009ApJ...691..866K}) it remains the best characterized extrasolar planet. The object has a mass $\mathcal{M} = 0.69 M_{\rm{Jup}}$, a radius $\mathcal{R} = 1.35 R_{\rm{Jup}}$ and a rotational (assumed synchronous with orbit) period of $\mathcal{T}_{\rm{HD209458b}} = 3.5$ days.

\paragraph{Momentum.} Because we are integrating a quasi-2D atmospheric shell, intended to be representative of a single pressure level, we apply impenetrable, stress-free boundary conditions to the velocity field. Thus, flows are free to develop without friction along the boundaries, although at initialization the planet is cast into solid-body rotation meaning $\mathbf{v} = 0$ at $t = 0$. 

\paragraph{Energy.} The temperature is initialized at $\bar{T} = 2000$K with $\delta T/\bar{T} = \delta \rho/\bar{\rho} = 0$. The atmosphere is kept stably-stratified by the imposition of a stable background temperature gradient (set to the adiabatic lapse rate $h = g/c_p$) similar to the implementation in other Boussinesq dynamo models with stable layers (e.g. \citet{2008PEPI..168..179S, 2010GeoRL..37.5201S, 2008Icar..196...16C}). Additionally, an azimuthally variable heat-flux is applied at the outer boundary, $r_2$, to account for the variable stellar irradiation (in practice it doesn't matter whether the variable heat flux is applied at the outer or the inner boundaries, since it is the horizontal temperature gradient that controls the flow). The functional form of the heat-flux is chosen to be that of the $\mathcal{F} \propto Y_1^1$ spherical harmonic, while the amplitude is taken to be a free parameter (see the discussion in the next section).

\paragraph{Magnetic Field.} Nominally, an electrical conductivity of $\sigma = 1$S/m, characteristic of $P \simeq 1$mbar, $T \simeq 2000$K is prescribed to the computational domain. Simultaneously, negligible electrical conductivity is assigned outside the computational domain. That is, $\sigma \simeq 0$ at $r < r_1$ and  $r > r_2$, meaning all of the current generated by the atmospheric flow is confined to the weather layer. This is in agreement with the approach of \citet{2010ApJ...719.1421P,2013ApJ...764..103R}, but in some contrast with the analytical work of \citet{2010ApJ...714L.238B, 2011ApJ...738....1B}, where the current is allowed to penetrate the convective interior of the planet. 

Because this work is primarily aimed at studying the upper atmosphere, the pressure-level of interest may reside above the atmospheric temperature inversion, characteristic of some Hot Jupiter atmospheres \citep{2007ApJ...668L.171B, 2009ApJ...699.1487S}. The presence of such an inversion may provide an electrically insulating layer\footnote{This idea is not new. The models of \citet{2010ApJ...714L.238B} as well as \cite{2011ApJ...738....1B} used the presence of such a layer to confine the current generated by zonal flows to the lower atmosphere and the interior of the planet.}, justifying $\sigma \simeq 0$ at $r < r_1$. The capability of electrical currents to penetrate the upper atmosphere and close in the magnetosphere is determined by the upper atmosphere's temperature structure and the abundance of Alkali metals at $P \lesssim 1$ mbar altitudes. Although a clear possibility, the physics of atmosphere-magnetosphere coupling is no-doubt complex and is beyond the scope of the present study. Consequently, we neglect it for the sake of simplicity. 

In all of our simulations, we initialize $\mathbf{J} = 0$ in the weather layer. However, the weather layer is still permeated by the background magnetic field, $\mathbf{B}_{\rm{dip}}$, presumed to be generated by dynamo action in the convective interior of the planet. This zero current field reads:
\begin{equation}
\label{Bdip}
\mathbf{B}_{\rm{dip}} = \nabla \times \left( k_m \frac{\sin (\theta)}{r^2} \hat{\mathbf{\phi}} \right),
\end{equation}
where $k_m$ is a constant that sets the surface field strength i.e. $|B_{\rm{dip}}| = k_m / \mathcal{R}^3$ at the equator. Within the context of the model, the background field is implemented by modifying the governing equations to account for a time-independent axially-dipolar external field. This occurs specifically in two terms. In the momentum equation (\ref{NavierStokes}), the Lorenz force $(\mathbf{J} \times \mathbf{B})/\bar{\rho}$ is replaced by $\mathbf{J} \times (\mathbf{B}+\mathbf{B}_{\rm{dip}})/\bar{\rho}$. In the magnetic induction equation (\ref{induction}), the induction term $\nabla \times (\mathbf{v} \times \mathbf{B})$ is replaced by $\nabla \times (\mathbf{v} \times (\mathbf{B}+\mathbf{B}_{\rm{dip}}))$. This method is similar to that used by \citet{1997Sci...276.1106S} and \citet{2012A&A...539A..78C} to implement external fields.

\subsection{Dimensionless Numbers}

Upon non-dimensionalization of the governing equations \citep{1999JCoPh.153...51K}, we obtain 6 dimensionless numbers that completely describe the system. They are the Ekman number $E$, the magnetic Rossby number $R_{\rm{o}}$, the magnetic Prandtl number $q$, the magnetic Rayleigh number $R_{\rm{th}}$ and the aspect ratio $\chi$ as well as an additional value, $\Gamma$, that parameterizes the incoming stellar flux. In terms of physical parameters, these quantities read:
\begin{eqnarray}
&E& \equiv \frac{\nu}{2 \Omega \mathcal{R}^2}, \ \ \ R_{\rm{o}} \equiv \frac{\eta}{2 \Omega \mathcal{R}^2}, \ \ \ \ \  \Gamma \equiv \frac{\mathcal{F}}{\kappa \bar{\rho} c_p h}, \nonumber \\
&q& \equiv \frac{\kappa}{\eta}, \ \ \ \ \ \ \  \  R_{\rm{th}} \equiv \frac{g h \mathcal{R}^2}{2 \bar{T} \Omega \eta}, \ \ \ \ \ \chi \equiv \frac{\mathcal{H}}{\mathcal{R}}.
\end{eqnarray}

The aspect ratio of the atmosphere we consider is $\chi = 7 \times 10^{-3}$. For numerical stability, the model requires the Ekman number to exceed a critical value (which we adopt in the simulations) of order $E_{\rm{crit}} \sim 10^{-5}$. If we consider the molecular viscosity of Hydrogen $\nu =  \bar{n} a \sqrt{3 k_{\rm{B}} \bar{T}/m_{\rm{H_2}}}/2$, where $\bar{n}$ is the number density and $a$ is the molecular cross-section, we obtain a hopelessly small  $E \sim 10^{-20} \ll E_{\rm{crit}}$ at mbar pressure. We note that the actual Ekman number is orders of magnitude higher, since on the global planetary scale, small-scale turbulence is a far more relevant source of viscosity than transfer of momentum at the molecular level \citep{1992phcl.book.....P}. Still, the true value of $E$ is probably much smaller than $E_{\rm{crit}}$.

For a given electrical conductivity (equivalently diffusivity), $R_{\rm{o}}$ and $R_{\rm{th}}$ simply encompass the physical units of the model. Specifically, for $\sigma = 1$S/m, $R_{\rm{o}} = 1.79 \times 10^{-6}$ and $R_{\rm{th}} = 1.38 \times 10^{9}$. Meanwhile, all of the information regarding the considered pressure level is provided by $q$. Taking $\sigma = 1$S/m as before, we obtain $q \simeq 6, 60$ and $600$, corresponding to $\tau_{\rm{N}} = 10^5, 10^4,$ and $10^3$ sec, appropriate for $P = 1, 0.1$ and $0.001$ bars respectively. Finally, as already mentioned above, we take $\Gamma$ to be an adjustable parameter, tuned to obtain the desired temperature gradients. It is important to note that our model differs somewhat from typical dynamo models in a sense that $\Gamma$, rather than $R_{\rm{th}}$, parameterizes the extent to which the system is driven.

Let us conclude the description of the numerical model with a brief discussion of its shortcomings as a guide for future work. First and foremost, the limitation of the atmosphere to a single scale-height may be prejudicial, as previous work has shown that including an extended vertical extent of the atmosphere is important to capture circulation features \citep{2011MNRAS.tmp..370H}. Second, while we have kept thermal and magnetic diffusivities uniform throughout the computational domain, it should be understood that in reality these values vary with temperature and pressure. Third, an implicit assumption of an infrared optical depth of order unity is rather crude at $\sim$ mbar pressure levels and should be lifted in more rigorous treatments of radiative transfer. Finally, as already mentioned above, the artificially enhanced viscosity inherent to our model almost certainly smoothes out smaller-scale flow to an unphysical extent.

\begin{figure}
\includegraphics[width=1\columnwidth]{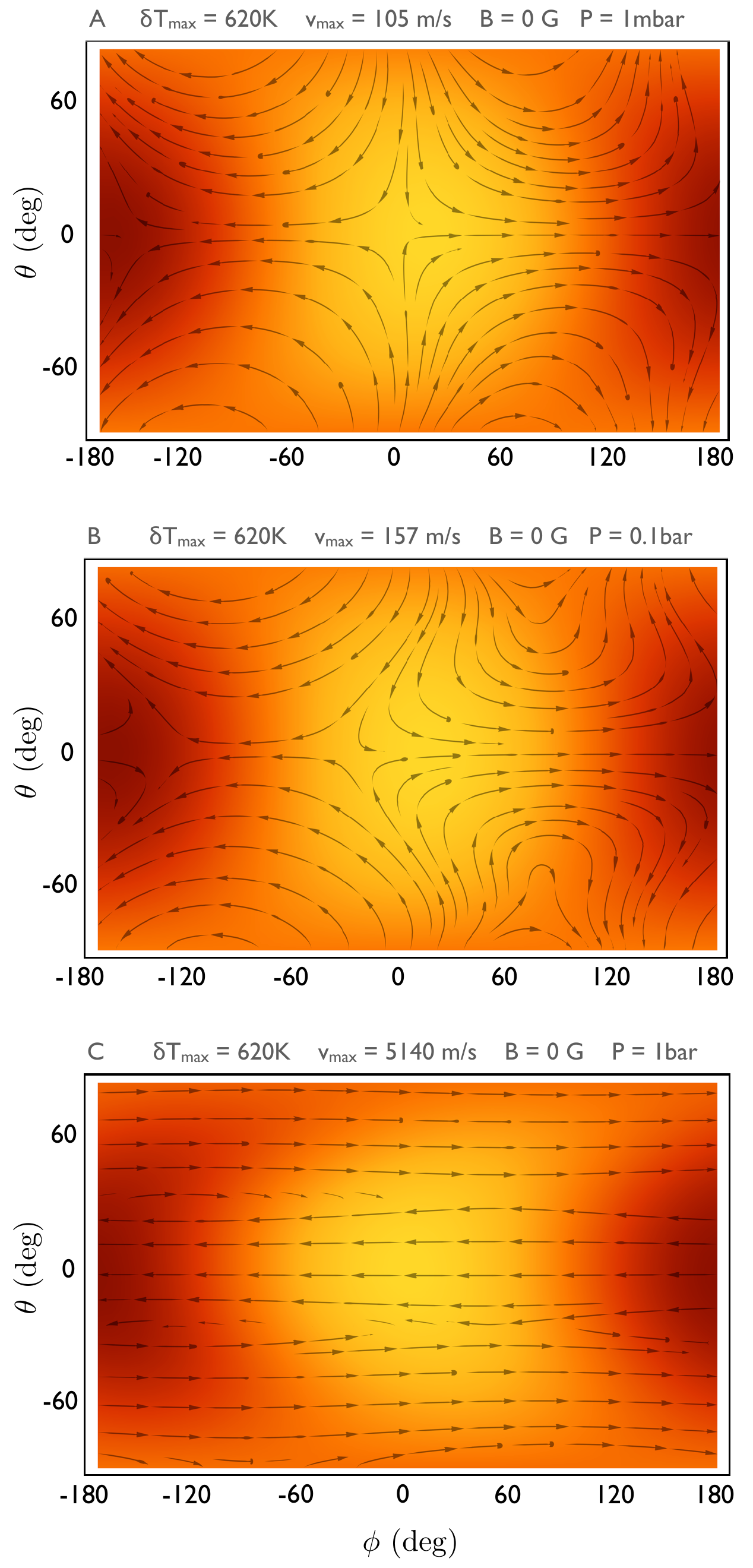}
\caption{Hydrodynamical simulations of the global circulation obtained using the numerical global circulation model of \citet{1999JCoPh.153...51K}. The arrows depict the currents of the flow and the color map is representative of the temperature structure. The background and initial atmospheric magnetic fields are set to zero, while the variable heat flux used to mimic insolation is tuned such that the maximal deviation of temperature from the background state is $\delta T_{\rm{max}}=620$K. The three panels of the Figure correspond to different pressure levels: $P = 1$ mbar (A), $P = 0.1$ bar (B) and $P = 1$ bar (C). Note that here, we have plotted the latitude, rather than colatitude used throughout the paper on the y-axis.}
\label{hydrosims}
\end{figure}

\subsection{Hydrodynamical Simulations}
Prior to performing unabridged MHD simulations, it is useful to first compare our model to previously published results. Accordingly, we begin by setting the strength of the background field to $\mathbf{B}_{\rm{dip}}=0$. Because the computational domain is initialized with a null current density, no magnetic fields are generated yielding purely hydrodynamic simulations. 

Horizontal slices of the atmosphere (through the center of the computational domain) in the cylindrical projection are shown in Figure (\ref{hydrosims}). The figures are centered on the substellar point and the background color shows the temperature distribution. The arrows denote the circulation vector field. Peak wind speeds as well as the maximal temperature deviations from $\bar{T} = 2000$K are labeled. 

The typically observed transition from zonal flows to dayside-nightside flows is observed as the pressure is decreased sequentially from $P = 1$ bars (Panel C of Figure \ref{hydrosims}) to $P = 0.001$ bars (Panel A of Figure \ref{hydrosims}). However, important differences exist in our results, contrasted against say, the results of \citet{2008ApJ...685.1324S}. The first important distinction is that in the zonally-dominated parameter regime, rather than developing a single broad jet, our model shows the development of three counter-rotating jets. 

This is a simple consequence of angular momentum conservation in the computational domain, and is not uncharacteristic of 2D models \citep{2011exop.book..471S}. Because of free-slip boundary conditions employed in our model, the atmosphere is not allowed to exchange angular momentum with the interior. As such, because the atmosphere is initialized in solid-body rotation, the development of any prograde jets must be accompanied by the development of retrograde jets. Because the retrograde jets reside at a high latitude $\theta_{\rm{ret}}$, whereas the prograde jets are essentially equatorial, angular momentum conservation requires them to be faster by a factor of $|\mathbf{v}_{\rm{ret}}/\mathbf{v}_{\rm{pro}}| \sim 1/\cos{\theta_{\rm{ret}}}$, as observed in the simulations. The angular momentum conserving 3D simulations of \citet{2011MNRAS.tmp..370H} exhibit similar behavior, although in their model the counter-rotating jets develop below the prograde ones, and are considerably slower because of the associated density enhancement. 

The second distinction of interest is the direction of the equatorial jet. While 3D GCMs consistently produce eastward equatorial jets, the equatorial jets in our hydrodynamical simulations are westward. This is not too surprising, as shallow-water and equivalent barotropic models are known to produce both eastward and westward equatorial jets, depending on the details of the simulation setup \citep{2011exop.book..471S,2011MNRAS.tmp..370H}. Consequently, this difference can probably be attributed to the limited vertical extent of our model.

The third important difference is the fact that flow velocities are not consistent along the pressure levels. In particular for the same values of $\delta T$, zonal flows have $\sim$ few km/s peak wind speeds, while the dayside-to-nightside flows are more than an order of magnitdue slower. This is largely a consequence of difference in the geometry of the circulation and the fact that viscosity enters as a significant member in the force balance for dayside-to-nightside flows. This can be seen by approximating $\nu \nabla^2 \sim \nu/ \mathcal{L}^2$ \citep{1992aitd.book.....H}, where $\mathcal{L}$ is a characteristic length scale associated with the curvature of the circulation. For zonal jets, $\mathcal{L}_{\rm{z}} \sim \mathcal{R}$ while for dayside-to-nightside circulation, the return flow is in part radial implying $\mathcal{L}_{\rm{dn}} \sim \mathcal{H}$. As will be shown in the following section, faster velocities can be attained by either increasing the aspect ratio of the atmosphere or by artificially enhancing the radiative heat flux, as a result of a linear proportionality between peak wind speeds and $\delta T$. An additional point of importance is that in typical 3D simulations, dayside-to-nightside circulation is nearly uniform over the terminator with the return flow residing at greater depth, while our results depict a partial return flow over the poles. This is again a consequence of the quasi-2D geometry of our model.

Although these quantitative distinctions are certainly worthy of attention, the typical features of the flow are approximately captured by our simplified model. Consequently, while being mindful of the model's limitation we do not view the quantitative dissimilarities as critical, as they are not central to the argument of the paper. After all, recall that we are primarily concerned with the possibility of a qualitative alteration of the dayside-to-nightside flow by magnetic effects. 

\section{Dragged Circulation in the Upper Atmosphere}

In the previous section, we performed baseline hydrodynamical simulations of atmospheric circulation at different pressure levels. In the following sections, we will focus primarily on the mbar pressure level, where the flow takes on a dayside-to-nightside character. As discussed above, in our simulations viscosity plays an important role in determining the flow velocities. Conceptually, the situation may be synonymous to simulations of invicid GCMs that parameterize the effect of magnetic coupling as Rayleigh drag \citep{2010ApJ...719.1421P,2012ApJ...750...96R}. In interest of understanding the dependence of the flow velocities on the magnitude of the dayside-to-nightside temperature gradient as well as the imposed frictional forces, in this section we shall develop a simple analytical model for dragged upper-atmospheric circulation and confirm it numerically.

\subsection{Analytical Theory}
Let us begin with estimation of characteristic timescales. In order to accomplish this, we first simplify the Lorentz and viscous forces to resemble the functional form of Rayleigh drag. Utilizing Ohm's law, we have\footnote{Although this approximation is often made, it is not necessarily sensible for systems where the magnetic Reynold's number, $Re_m \equiv v \mathcal{L}/\eta \gg 1$. Adopting $v \sim$ km/s and $\mathcal{L} \sim \mathcal{H}$, we obtain $Re_m \sim 10^3$, placing the magnetic drag approximation on shaky footing. Further, the electric field in Ohm's law can only be neglected when the radial current is negligible. } :
\begin{eqnarray}
\label{drag}
\frac{\mathbf{J} \times \mathbf{B}}{\bar{\rho}} &\sim& \frac{\sigma (\mathbf{v} \times \mathbf{B}_{\rm{dip}}) \times \mathbf{B}_{\rm{dip}} }{\bar{\rho}} \sim - \left( \frac{\sigma k_m^2}{\mathcal{R}^6\bar{\rho}} \right) \mathbf{v} = - \frac{\mathbf{v}}{\tau_{\rm{L}}} \nonumber \\
\nu \nabla^2 \mathbf{v} &\sim& - \left( 3 \frac{\nu}{\mathcal{R}^2} \right) \mathbf{v} = - \frac{\mathbf{v}}{\tau_{\nu}}
\end{eqnarray}
This allows us to rewrite equation (\ref{NavierStokes}) in a simpler form:
\begin{eqnarray}
\frac{D \mathbf{v}}{D t} = - 2 \mathbf{\Omega} \times \mathbf{v}  - \frac{\nabla \mathcal{P}}{\bar{ \rho} } + \frac{\delta \rho}{\bar{\rho}} \mathbf{g} - \frac{\mathbf{v}}{\tau_{f}},
\end{eqnarray}
where $\tau_{f} = (1/\tau_{\rm{L}} + 1/\tau_{\nu})^{-1}$. Taking $|B_{\rm{dip}}| = 1$ G, the characteristic timescales are: $\tau_{L} \sim 10^{3}$ sec and $\tau_{\nu} \sim 10^{5}$ sec. Other relevant timescales in the problem is the rotational (Coriolis) timescale $\tau_{\Omega} \sim 2 \pi / \Omega \sim 10^{5}$ sec, radiative timescale $\tau_{\rm{N}} \sim 10^{3}$ sec and the advective timescale $\tau_{\rm{adv}} \sim \mathcal{R}/v  \sim 10^{5}$ sec. 

There exists a clear separation of timescales in the system. As a result, upon including the parameterized Lorentz force into the equations of motion, the inertial and Coriolis terms can be dropped (the viscous term can be dropped as well, although this simplification is unnecessary). The removal of the inertial terms implies a steady-state solution. The removal of the Coriolis term creates a symmetry characterized by an axis that intersects the sub-solar and anti-solar points. Taking advantage of this symmetry, we orient the polar axis of the coordinate system such that it intersects the sub-solar point. Upon doing so, we can specify a null azimuthal velocity and drop all azimuthal derivatives in the equations of motion. In a local cartesian reference frame, this leaves us with horizontal ($\hat{\mathbf{y}}$) and vertical ($\hat{\mathbf{z}}$) momentum equations, where the latter simplifies to the equation of hydrostatic balance:
\begin{eqnarray}
\label{carteqns}
\frac{1}{\bar{\rho}}\frac{\partial \mathcal{P}}{\partial y} = - \frac{v_y}{\tau_f} \nonumber \\
\frac{1}{\bar{\rho}}\frac{\partial \mathcal{P}}{\partial z} = g \frac{\delta T}{\bar{T}}
\end{eqnarray}

Following \citet{1977JAtS...34..263S} and \citet{1980JAtS...37..515H}, we shall adopt a Newtonian energy equation with implicit stable stratification (recall that we have set the potential temperature gradient to $h = g/c_p$): 
\begin{equation}
\label{newtenergy}
\xi v_z \left( \frac{g}{c_p} \right) = - \frac{\delta T - \delta T_{\rm{rad}}}{\tau_{\rm{N}}}.
\end{equation}
Here, $\xi \geqslant 1$ is a constant that parameterizes lateral heat advection and $\delta T_{\rm{rad}}$ is a purely radiative perturbation to the background state, $\bar{T}$. In other words, as the damping of the circulation is strengthened and $v \rightarrow 0$, $\delta T \rightarrow \delta T_{\rm{rad}}$. 

Retaining the incompressibility condition (\ref{continuity}), we introduce a stream-function $\Psi$, defined through $\mathbf{v} = \nabla \times \Psi$ \citep{1959flme.book.....L}. Taking a partial derivative of the $y$-momentum equation with respect to $z$ and of the $z$-momentum equation with respect to $y$, we obtain
\begin{equation}
\label{eq1}
-\frac{1}{\tau_{f}} \frac{\partial^2 \Psi}{\partial z^2} = \frac{g}{\bar{T}} \frac{\partial (\delta T)}{\partial y}.
\end{equation}

\begin{figure*}
\includegraphics[width=1\textwidth]{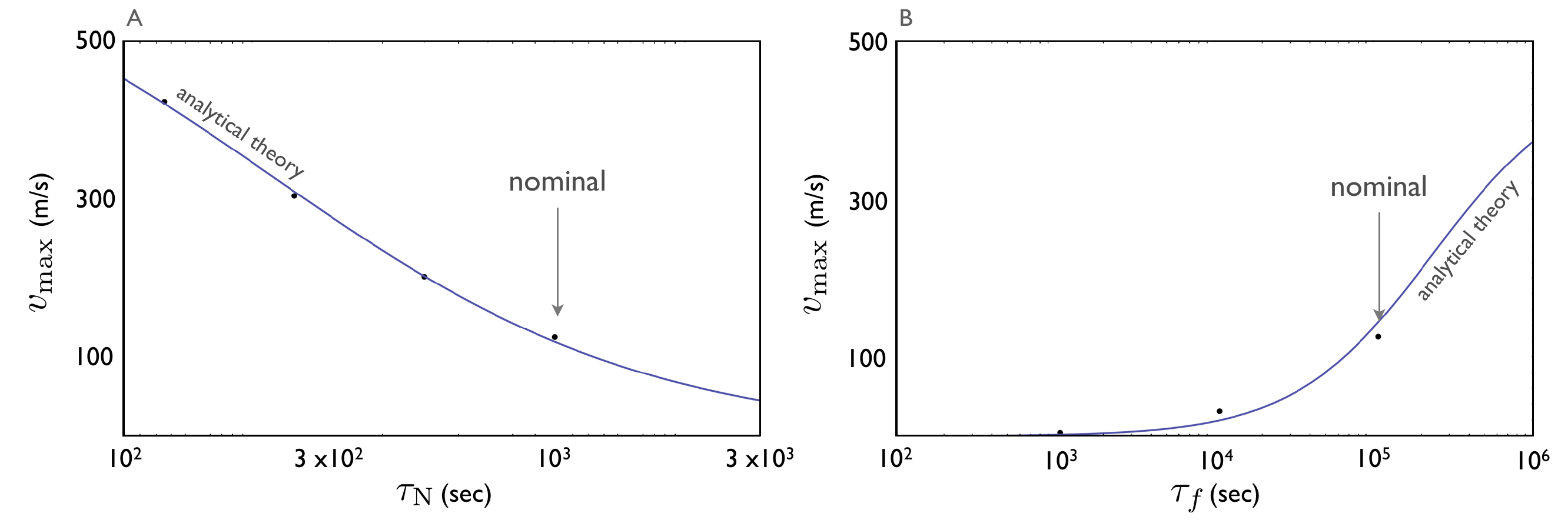}
\caption{Dependence of peak wind speeds on the radiative and drag timescales obtained within the context of dragged hydrodynamical solutions. The black points represent the results of numerical experiments, where enhanced viscosity is used to mimic the effects of the magnetic field, while the curves represent the analytical solution derived from equations (\ref{eq5}) and (\ref{eq7}). The values $\xi = 8630$ and $\zeta = 0.16$ have been adopted for the analytical solution. Note that the fact that $\xi \gg 1$ is simply a consequence of the fact that lateral advection of heat completely dominates over vertical advection of heat as the transport mechanism of importance. That is to say that in the numerical solutions, $v_{z} (\partial T/\partial z) \ll v_{y} (\partial T/\partial y)$ Nominal values of $\tau_{f}$ and $\tau_{\rm{N}}$ are adopted in panels A and B respectively.}
\label{draggedsims}
\end{figure*}

Taking a derivative of equation (\ref{eq1}) with respect to y and switching the order of partial differentiation yields:
\begin{equation}
\label{eq2}
- \frac{1}{\tau_{f}} \frac{\partial}{\partial z} \left(  \frac{\partial^2 \Psi}{\partial y \partial z}  \right)= \frac{g}{\bar{T}} \frac{\partial^2 (\delta T)}{\partial y^2}.
\end{equation}
Meanwhile, differentiating the Newtonian cooling equation (\ref{newtenergy}) with respect to z gives:
\begin{equation}
\label{eq3}
\left(  \frac{\partial^2 \Psi}{\partial y \partial z}  \right) =  \frac{c_p}{g \xi} \frac{\partial}{\partial z} \left( \frac{\delta T - \delta T_{\rm{rad}}}{\tau_{\rm{N}}} \right),
\end{equation}
allowing us to eliminate $\partial^2 \Psi/\partial y \partial z$:
\begin{equation}
\label{eq4}
\frac{\partial^2 (\delta T_{\rm{rad}})}{\partial z^2} -  \frac{\partial^2 (\delta T)}{\partial z^2} = \xi \frac{ \tau_{f} \tau_{\rm{N}} g^2}{c_p \bar{T}} \frac{\partial^2 (\delta T)}{\partial y^2}.
\end{equation}
Note that $g^2/c_p \bar{T}$ is simply the square of the Brunt-Vaisala frequency for an isothermal atmosphere.

Equation (\ref{eq4}) admits the trial solutions
\begin{eqnarray}
\label{eq5}
&\delta& T_{\rm{rad}} = (\delta T_{\rm{rad}}^0) \cos\left( \frac{y}{\mathcal{R}} \right)  \sin \left(\frac{\pi  z }{\mathcal{\mathcal{H}}} \right) \nonumber \\
&\delta& T = (\delta T^0) \cos\left( \frac{y}{\mathcal{R}} \right)  \sin \left(\frac{\pi  z }{\mathcal{\mathcal{H}}} \right),
\end{eqnarray}
where $(\delta T_{\rm{rad}}^0)$ and $(\delta T^0)$ are constants that represent the maximal deviations in the respective quantities from the background state. Note that $(\delta T_{\rm{rad}}^0)$ is a parameter inherent to the model rather than a variable. Upon substitution of the above solutions (\ref{eq5}) into equation (\ref{eq4}), we obtain a relationship between $(\delta T^0)$ and $(\delta T_{\rm{rad}}^0)$:
\begin{equation}
\label{eq6}
(\delta T^0) = (\delta T_{\rm{rad}}^0) \left[ 1 + \xi \zeta \frac{ \tau_{f} \tau_{\rm{N}} g^2}{\pi^2 c_p \bar{T}} \left( \frac{\mathcal{H}}{\mathcal{R}} \right)^2 \right]^{-1}.
\end{equation}
In the above equation, $\zeta$ is an empirical factor that has been introduced to account for the approximations inherent to equations (\ref{drag}). 

With an analytical solution for the temperature perturbation in hand, we substitute equations (\ref{eq5}) into equation (\ref{eq1}) and integrate twice to obtain:
\begin{equation}
\label{eq7}
\Psi = \left[ \frac{g \zeta \tau_{f} \mathcal{H}^2 (\delta T^0) }{\pi^2 \bar{T} \mathcal{R}} \right] \sin\left( \frac{y}{\mathcal{R}} \right) \sin \left(\frac{\pi z}{\mathcal{H}} \right) \hat{\mathbf{x}}.
\end{equation}
This solution implies the same functional form for laterally-averaged heat transport in the vertical and horizontal advection terms in the energy equation (\ref{heatequation}), lending some support for the approximation inherent to equation (\ref{newtenergy}). Once the stream function is obtained, the maximal horizontal and vertical velocities are given by:
\begin{equation}
\label{peakwind}
v_y^{\rm{max}} = \Psi_0 \frac{\pi}{\mathcal{H}} \ \ \ \ \ v_z^{\rm{max}} = \Psi_0 \frac{1}{\mathcal{R}},
\end{equation}
where $\Psi_0$ is the term in square brackets in (\ref{eq7}). Note that the above theory automatically implies a quasi-2D flow since $v_z^{\rm{max}}/v_y^{\rm{max}} \sim (\mathcal{H}/\mathcal{R}) \ll 1.$ 

\subsection{Numerical Experiments}
With a simple analytical theory at hand, we performed a series of numerical simulations, varying the radiative and drag timescales in the ranges $10^2 < \tau_{N} < 10^3$ sec and $10^3< \tau_{f} < 10^5$ sec. Although we observed a considerable variability in the wind speeds and dayside-to-nightside temperature differences in our simulations, the nature of the flow was largely the same as that seen in panel A of Figure (\ref{hydrosims}) across the runs. The peak wind speeds obtained in the simulations as functions of $\tau_{N}$ and $\tau_{f}$ are presented as black dots in Figure (\ref{draggedsims}).

In addition to the simulation results, Figure (\ref{draggedsims}) shows $v_y^{\rm{max}}$ given by equation (\ref{peakwind}) for the same parameters. As can be seen from the figures, the scaling law inherent to equation (\ref{eq7}) matches the numerical experiments quite well. Extrapolating towards larger values of $\tau_{f}$, it can be inferred that our simulations would have produced peak wind speeds of order $\sim$ km/s if not for the numerical requirements of enhanced viscosity. However, it can also be expected that the character of the flow would also change qualitatively with diminishing viscosity, as the force balance shifts away from that implied by equations (\ref{drag})\footnote{In fact, even for the nominal simulation shown in panel A of Figure (\ref{hydrosims}), the flow patterns do not exactly follow the analytical streamfunction (\ref{eq7}).}.

It should be kept in mind that the adjustable parameters $\xi$ and $\zeta$ were fit to the data. Moreover, the value of $(\delta T_{\rm{rad}}^0) = 3360$K\footnote{Note that strictly speaking, this is an unphysical value, that highlights the limitations of our model.} was chosen by running a simulation where viscous forces completely dominated the force balance ensuring $\mathbf{v} = 0$. In other words, the quantitative agreement seen in Figure (\ref{draggedsims}) is a consequence of the fact that the adjustable parameters of the analytical solution have been fit to the numerical data, but the fact that the functional form of the analytical model conforms with numerical experiments suggest that the stream-function (\ref{eq7}) captures the main features of dragged upper-atmospheric circulation on Hot Jupiters.

\section{Magnetically Controlled Circulation}
In the last section, we examined the extent to which dayside-nightside flow can be damped by imposing a drag. However, both the analytical theory and numerical experiments showed that the qualitative character of the circulation remained largely unchanged. As already mentioned in the introduction, the rough consistency of the flow patterns across a range of characteristic drag timescales is in broad agreement with the results of ``primitive" 3D GCMs \citep{2010ApJ...719.1421P,2012ApJ...750...96R}. In this section, we challenge this assertion with MHD calculations. 

\subsection{Theoretical Arguments}

With the exception of a rather limited number of problems, self-consistent magneto-hydrodynamic solutions can only be attained with the aid of numerical simulations. However, for the system at hand, the qualitative effect of magnetic induction can be understood from simple theoretical considerations. As we already argued, the two characteristic states of Hot Jupiter atmospheric circulation are a zonally-dominated state and a meridionally-dominated state \citep{2011exop.book..471S}. Whether or not a given configuration will be significantly affected by the introduction of the magnetic field can be established by analyzing its stability. More specifically, we can work within a purely kinematic (rather than dynamic) framework to understand if the Lorentz force acts to perturb the flow away from its hydrodynamic counterpart or simply damps the circulation. 

\paragraph{Zonal Flows.} Although not directly applicable, recent studies of Ohmic dissipation that arises from zonal flows performed by \citet{2008Icar..196..653L} (within the context of solar system gas giants) and by \citet{2010ApJ...714L.238B} as well as \citet{2012ApJ...745..138M} (within the context of Hot Jupiters) have already produced some results on a related problem. Here we work in the same spirit as these studies and prescribe the following functional form to the zonal flow to approximately represent three jets, such as those shown in panel C of Figure (\ref{hydrosims}):
\begin{equation}
\label{vzonal}
\tilde{\mathbf{v}} = \tilde{v}_0 \sin (3 \theta) \hat{\mathbf{\phi}},
\end{equation}
where $\tilde{v}_0$ is a negative constant, whose magnitude corresponds to the peak wind speed. This prescription trivially satisfies the continuity equation (\ref{continuity}), although we note that a more realistic zonal flow should also exhibit differential rotation.

The interaction between this flow and the background magnetic field (\ref{Bdip}) will induce a field $\mathbf{B}_{\rm{ind}}$ in the atmosphere. Because $\mathbf{B}_{\rm{dip}}$ is entirely poloidal, and $\tilde{\mathbf{v}}$ is strictly toroidal, $\mathbf{B}_{\rm{ind}}$ will also be strictly toroidal \citep{1978mfge.book.....M}. As can be readily deduced from equation (\ref{induction}), this means that $\mathbf{B}_{\rm{ind}}$ cannot interact with $\tilde{\mathbf{v}}$ to further induce new field unless $\tilde{\mathbf{v}}$ deviates from a purely zonal flow. As a result, in steady state, the induction equation reads:
\begin{equation}
\label{ssinduction}
-\eta \nabla^2 \mathbf{B}_{\rm{ind}} = \nabla \times ( \tilde{v} \times \mathbf{B}_{\rm{dip}} ) = -\frac{6 k_m \tilde{v}_0 }{r^4} \cos (\theta) \sin (\theta) \hat{\mathbf{\phi}}.
\end{equation}
It is noteworthy that had we chosen to represent a single broad jet (such as that seen in most 3D simulations \citep{2008ApJ...685.1324S, 2010ApJ...713.1174M, 2013ApJ...764..103R}) by setting $\tilde{v} \propto \sin (\theta)$ (in this case $\tilde{v}_0$ is positive), equation (\ref{ssinduction}) would have looked the same, with the exception of the coefficient on the RHS, which would have been $2$ instead of $6$. As a result, it should be kept in mind that the following kinematic solution applies to the case of a single jet as well.

\begin{figure}
\includegraphics[width=1\columnwidth]{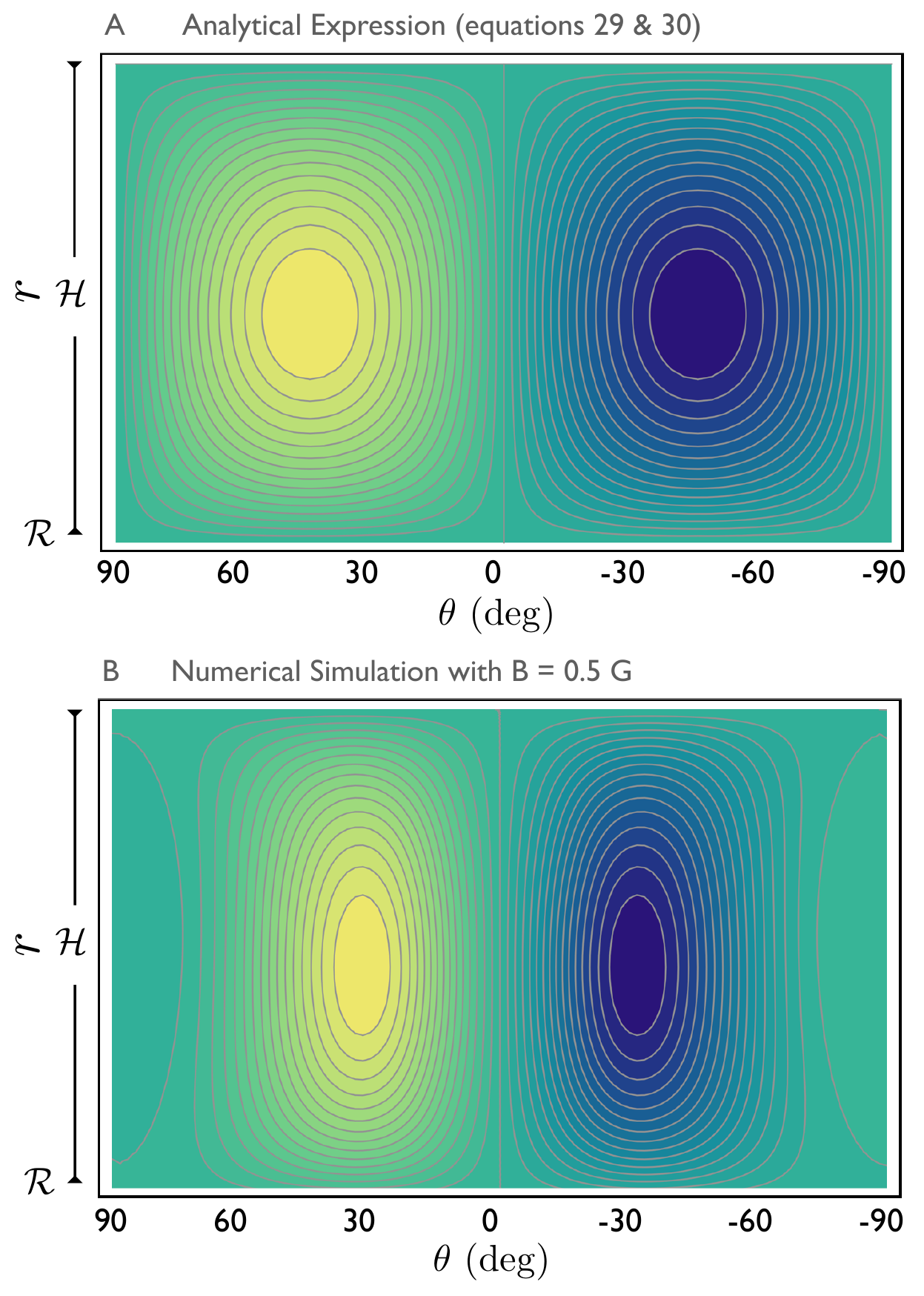}
\caption{Toroidal magnetic fields induced in the atmospheric shell. Panel A represents the kinematic analytical solution obtained through equations (\ref{Bindtheta}) and (\ref{BindthetaA}), while panel B depicts a result obtained from a dynamic numerical simulation, also shown in panel C of Figure (\ref{mhdsims}). Green colors correspond to eastward (positive) fields, and the converse is true for blue colors. The contour lines depict the associated electrical currents. The maximal induced field strengths are $\rm{max}(B_{\rm{ind}}) = 0.64$ G and $\rm{max}(B_{\rm{ind}}) = 0.52$ G corresponding to the analytical solution (panel A) and the numerical solution (panel B) respectively. Because the Lorentz force associated with this induced field acts in the same sense as the flow itself, it can only act to accelerate/decelerate the jets but not alter their directions.}
\label{inducedfield}
\end{figure}

The angular part of equation (\ref{ssinduction}) is satisfied by the expression
\begin{equation}
\label{Bindtheta}
\mathbf{B}_{\rm{ind}} = \mathcal{A}(r) \cos (\theta) \sin (\theta) \hat{\mathbf{\phi}},
\end{equation}
where $\mathcal{A}(r)$ is a yet undefined function. This form ensures that the meridional component of the induced current vanishes at the poles. Meanwhile, the radial impenetrability of the boundaries requires $\mathcal{A}(\mathcal{R}) = \mathcal{A}(\mathcal{R}+ \mathcal{H}) = 0$ as dictated by equation (\ref{J}). With these boundary conditions, equation (\ref{ssinduction}) can be solved to yield
\begin{eqnarray}
\label{BindthetaA}
\mathcal{A}(r) &=&  3 k_m \tilde{v}_0 (\mathcal{R}-r) (\mathcal{H}^4+5 \mathcal{H}^3 \mathcal{R}+10 \mathcal{H}^2 \mathcal{R}^2 \nonumber \\
&+&10 \mathcal{H} \mathcal{R}^3+4 \mathcal{R}^4-\mathcal{R}^3 r-\mathcal{R}^2 r^2-\mathcal{R} r^3-r^4 ) / ( 2 \eta r^3 \nonumber \\
&\times& (\mathcal{H}^4+5 \mathcal{H}^3 \mathcal{R}+10 \mathcal{H}^2 \mathcal{R}^2+10 \mathcal{H} \mathcal{R}^3+5 \mathcal{R}^4 )).
\end{eqnarray}
The induced field and the associated electrical current are shown in panel A of Figure (\ref{inducedfield}).

The Lorentz force that arises from the interactions between $\mathbf{B}_{\rm{ind}}$ and $\mathbf{B}_{\rm{dip}}$ takes the form
\begin{eqnarray}
\mathbf{F}_{\rm{L}} &=& \frac{(\nabla \times \mathbf{B}_{\rm{ind}}) \times \mathbf{B}_{\rm{dip}}}{\rho \mu_0} = \left( \frac{ \sigma k_m^2}{r^6 \rho} \right) \tilde{v}_0 \hat{\mathbf{\phi}} \nonumber \\ 
&\times& (3 \sin (\theta ) (-11 \mathcal{R} (\mathcal{H}+\mathcal{R}) (\mathcal{H}+2 \mathcal{R}) (\mathcal{H}^2+2 \mathcal{H} \mathcal{R} \nonumber \\
&+& 2  \mathcal{R}^2)+9 r (\mathcal{H}^4+5 \mathcal{H}^3 \mathcal{R}+10 \mathcal{H}^2 \mathcal{R}^2+10 \mathcal{H} \mathcal{R}^3 \nonumber \\
&+& 5  \mathcal{R}^4)-r^5)+\sin (3 \theta ) (-7 \mathcal{R} (\mathcal{H}+\mathcal{R}) (\mathcal{H}+2 \mathcal{R}) \nonumber \\
&\times& (\mathcal{H}^2 + 2 \mathcal{H} \mathcal{R}+2 \mathcal{R}^2)+5 r (\mathcal{H}^4+5 \mathcal{H}^3 \mathcal{R} \nonumber \\
&+&10 \mathcal{H}^2 \mathcal{R}^2+10 \mathcal{H} \mathcal{R}^3+5 \mathcal{R}^4)+3 r^5))/(8 r \nonumber \\
&\times&   (\mathcal{H}^4+5 \mathcal{H}^3 \mathcal{R}+10 \mathcal{H}^2 \mathcal{R}^2+10 \mathcal{H} \mathcal{R}^3+5 \mathcal{R}^4)).
\end{eqnarray}
Because the Lorentz force acts in the same sense as the flow itself (that is, $\mathbf{F}_{\rm{L}} \times \tilde{\mathbf{v}} = 0$), it can only act to accelerate/decelerate the jets but not alter their directions. Indeed, the functional form of $F_{\rm{L}}$ is that of a Rayleigh drag (equation \ref{drag}), however the characteristic timescale is non-uniform in latitude and radius i.e. $\tau_{\rm{L}} = f(r,\theta)$. The non-uniformity we derive here should not be confused with the variability in $\mathbf{F}_{\rm{L}}$ that can arise from the spatial dependence of the electrical conductivity (see \citet{2013ApJ...764..103R}).

It is noteworthy that the radial dependence of $\mathbf{F}_{\rm{L}}$ can give rise to differential rotation. However, this is not particularly important, since in some similarity with the above discussion, differential rotation will only induce toroidal fields through the $\omega$-effect \citep{1978mfge.book.....M} and will therefore only change the solution obtained here on a detailed level (i.e. the added dependence of $\tilde{\mathbf{B}}$ on $r$ will subtly modify the function $\mathcal{A}(r)$). In other words, a differentially rotating zonal flow still results in a purely toroidal induced field.  

For a sensible comparison with previous works (e.g. \citep{2010ApJ...719.1421P,2012ApJ...754L...9M,2012ApJ...750...96R}) and the simple theory presented in the previous section, it is instructive to evaluate the maximal magnitude of $\mathbf{F}_{\rm{L}}$, which corresponds to the upper and lower boundaries of the domain in question\footnote{Although the Lorentz force is equal and opposite at $r = \mathcal{R}$ and $r = \mathcal{R} + \mathcal{H}$, its radial distribution is such that its vertically integrated value acts to oppose the flow on average.} i.e. $r = \mathcal{R}, r = \mathcal{R} + \mathcal{H}$. To leading order in $\chi$, the expression reads:
\begin{equation}
\label{maxFL}
\rm{max}(\mathbf{F}_{\mathrm{L}}) \simeq - 6 \left( \frac{\sigma \mathit{k}_m^2}{\mathcal{R}^6 \rho} \right) \tilde{v}_0  \chi \cos(\theta)^2 \sin (\theta) \hat{\mathbf{\phi}}.
\end{equation}

From this expression it is clear that $\mathbf{F}_{\rm{L}}$ acts primarily in the mid-latiudes rather than the equator. As a result, the damping of the jets is latitudinally differential, meaning that even if the flow is initially composed of multiple bands (as we consider here), it will approach a single equatorial jet as the conductivity and/or the magnetic field is increased. Furthermore, recall that the functional form of equation (\ref{maxFL}) is also valid in the case of a single jet. Qualitatively, this seems to imply that the Lorentz force acts to collimate the jet towards the equator. Such an effect is sensible given that the radial component of the field is stronger as one approaches the pole for a simple dipole. However, it should also be kept in mind that a true planetary magnetic field might be more complicated, leading to further lack of triviality in the circulation.

\paragraph{Dayside-to-Nightside Flows.} Let us now consider the more topologically complex interaction between meridional flows and a spin-pole aligned dipole magnetic field. As in section (3.1) we shall work in a coordinate frame where the polar axis intersects the sub-solar point and is directed at the host star. Unlike the case of zonal circulation, this configuration has no exploitable symmetry. Consequently, a simple solution to the steady-state induction equation (\ref{ssinduction}) is difficult, if not impossible, to obtain. We shall therefore make substantial simplifications.

In our prescription for the velocity field, we neglect radial flow altogether (thereby violating continuity) and adopt an expression similar to equation (\ref{vzonal}):
\begin{equation}
\label{vdaynight}
\tilde{\mathbf{v}} = \tilde{v}_0 \sin (\theta) \hat{\mathbf{\theta}}.
\end{equation}

Because of our choice of coordinate system, equation (\ref{Bdip}) cannot be used directly. However, keeping in mind that the background dipole field originates in a much deeper region of the planet than the atmosphere, we can write down the magnetic field in a current-free representation \citep{1998clel.book.....J}: 
\begin{equation}
\mathbf{B}_{\rm{dip}} = - \nabla  \Upsilon = - \nabla \left( k_m \frac{\sin(\theta) \cos(\phi)}{r^2} \right).
\end{equation}

As in \citet{2010ApJ...714L.238B}, we assume that the induction term is dominated by the interaction with the background field, rather than the induced field: $(\tilde{\mathbf{v}} \times \mathbf{B}) \simeq (\tilde{\mathbf{v}} \times \mathbf{B}_{\rm{dip}})$. Upon making this simplification and uncurling equation (\ref{induction}), the steady state induction equation reduces to Ohm's law:
\begin{equation}
\mathbf{J} = \sigma \left( \tilde{\mathbf{v}} \times \nabla  \Upsilon - \nabla \Phi \right), 
\end{equation}
where $\nabla \Phi$ is the electric field.

Because the current is necessarily divergence-free, the scalar potential $\Phi$ can be obtained from the following equation:
\begin{equation}
\nabla^2 \Phi = \nabla \cdot \left(\tilde{\mathbf{v}} \times \nabla  \Upsilon \right) = \frac{k_m \tilde{v}_0}{r^3} \sin (\theta) \sin(\phi). 
\end{equation}
It can be easily checked that the angular part of this relationship is satisfied by:
\begin{equation}
\Phi = \mathcal{A}(r) \sin (\theta) \sin(\phi).
\end{equation}

As before, confining the current to the atmosphere implies the boundary conditions: $k_m \tilde{v}_0 = \mathcal{R}^3 \mathcal{A}(\mathcal{R}) = (\mathcal{R}+\mathcal{H})^3 \mathcal{A}(\mathcal{R}+\mathcal{H})$. In turn, the radial part of the solution reads:
\begin{eqnarray}
\mathcal{A}(r) &=& k_m \tilde{v}_0 ( (-\mathcal{H} (\mathcal{H}^2+3 \mathcal{H} \mathcal{R}+3 \mathcal{R}^2) (\log (r)+2)  \nonumber \\
&+& \log (\mathcal{R}) (\mathcal{H}^3+3 \mathcal{H}^2  \mathcal{R}+3 \mathcal{H} \mathcal{R}^2+\mathcal{R}^3+2 r^3)  \nonumber \\
&-&(\mathcal{R}^3+2 r^3) \log (\mathcal{H}+\mathcal{R}) ) )/(3 \mathcal{H} r^2 \nonumber \\
&\times& (\mathcal{H}^2+3 \mathcal{H} \mathcal{R}+3 \mathcal{R}^2)).
\end{eqnarray}
The induced current density can now be obtained through Ohm's law. 

The Lorentz force can be approximated as originating from the interactions between the induced current and the background magnetic field. The resulting expression is quite cumbersome. However, all of the important features of $\mathbf{F}_{\rm{L}}$ can be seen by evaluating it at the center of the dynamic domain. To leading order in $\chi$, the expression takes the form:
\begin{eqnarray}
&\mathbf{F}&_{\rm{L}}|_{(r = \mathcal{R}+\mathcal{H}/2)} \simeq \left(  \frac{\sigma  k_m^2 }{\mathcal{R}^6 \rho} \right ) \tilde{v}_0  \nonumber \\
&\times& \Big[ \hat{r} \cos (\theta ) (1 -2 \sin ^2(\theta ) \cos (2 \phi )+ \cos (2 \theta ) )/2 \nonumber \\
&+& \hat{\mathbf{\theta}}  (\sin (3 \theta )-\sin (\theta )) \cos ^2(\phi ) \nonumber \\
&-& 2 \hat{\mathbf{\phi}} \sin (\theta ) \cos (\theta ) \sin (\phi ) \cos (\phi ) \Big].
\end{eqnarray}
A similar evaluation of $\mathbf{F}_{\rm{L}}$ at $r = \mathcal{R}$ and $r = \mathcal{R}+\mathcal{H}$ shows that the $\hat{r}$ and $\hat{\phi}$ components of the force do not change significantly with radius, although the $\hat{\theta}$ component does.

Indeed, the Lorentz force that arises from the interactions between the dayside-to-nightside circulation and the background magnetic field does not only oppose the flow. Instead, it acts to introduce both radial and zonal components to the circulation. Importantly, the typical magnitude of the zonal component of $\mathbf{F}_{\rm{L}}$ is commensurate with the meridional component (although of course their spatial dependence is different). As argued in section (3.1), the characteristic timescale associated with the Lorentz force is comparable to the radiative timescale at mbar pressures and is generally shorter than that, corresponding to other relevant forces. This means that the force-balance implied by equations (\ref{carteqns}) is in essence not relevant to circulation on hot planetary atmospheres. 

In summary, we conclude that dayside-to-nightside flow is unstable to perturbations arising from the Lorentz force. Consequently, we expect that the upper atmospheric circulation will change qualitatively once a substantial magnetic field is introduced into the system. We now turn our attention to numerical MHD simulations with the aim of testing this presumption and quantifying the dynamical state of magnetized upper atmospheres of Hot Jupters.

\begin{figure}
\includegraphics[width=1\columnwidth]{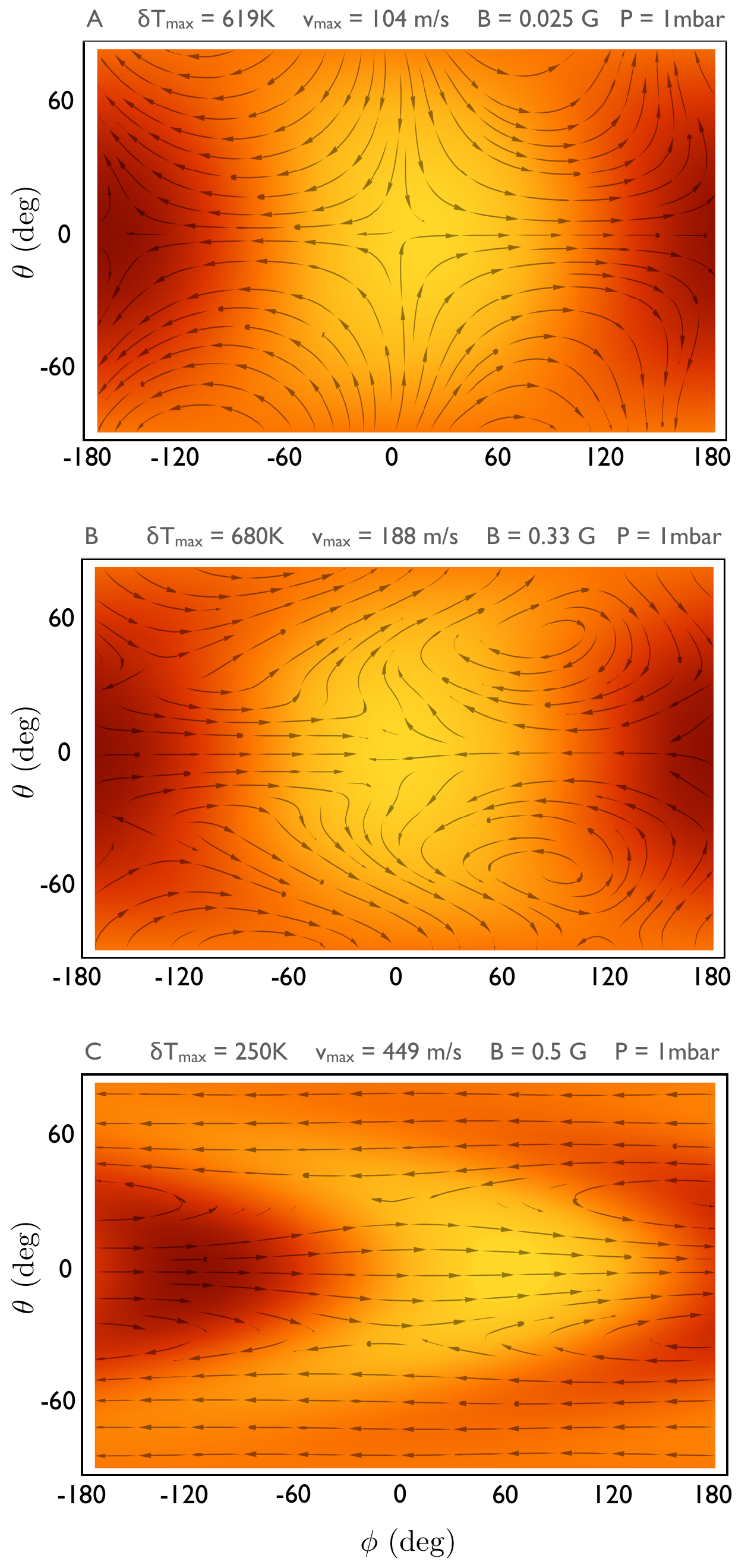}
\caption{Magnetohydrodynamical simulations of the global circulation performed using the numerical model of \citet{1999JCoPh.153...51K}. The arrows depict the currents of the flow and the color map is representative of the temperature structure. All panels correspond to the same pressure level, namely $P = 1$ mbar. However, the background magnetic field is gradually increased from panel A to panel C. Clearly, the qualitative character of the flow changes substantially once the field is increased above $B \simeq 0.3$ G (Panel B). The fact that the flow transitions to a globally zonal state at a value of background field strength that is considerably smaller than the inferred field surface strengths of Hot Jupiters suggests that dayside-to-nightside flows exist only on planets occupying the cooler end of close-in orbital distribution. For reference, the amplitudes of the axisymmetric component of the induced field corresponding to panels A, B, and C are $\rm{max}(B_{\rm{ind}}) = 0.075, 0.35$, and $0.52$ G respectively. Note that here, we have plotted the latitude, rather than colatitude used throughout the paper on the y-axis.}
\label{mhdsims}
\end{figure}

\subsection{Magnetohydrodynamic Simulations}

The hydrodynamic simulation parameters are chosen as described in section (2), corresponding to the $P = 1$mbar pressure level (i.e. $\tau_{\rm{L}} = 10^3$ sec.). We start out with the equilibrated hydrodynamic flow shown in panel A of Figure (\ref{hydrosims}) and introduce a weak pole-aligned dipole magnetic field into the system. Upon equilibration, we take the approach of sequentially increasing the magnitude of $B_{\rm{dip}}$. At each step, we allow the flow to reach a steady state before increasing the field strength further. We have checked that the flows obtained by successive enhancement of $B_{\rm{dip}}$ are identical to those obtained by initializing the atmosphere in solid-body rotation with a given value of $B_{\rm{dip}}$. Consequently, in agreement with \citet{2012arXiv1208.0126L}, we conclude that the obtained flows are insensitive to initial conditions. 

The panels of Figure (\ref{mhdsims}) show the upper atmospheric circulation for a series of magnetic field strengths. From this series of results, a clear pattern emerges: as the magnitude of $B_{\rm{dip}}$ is increased, the flow takes on an exclusively zonal character. Specifically, it is clear that the circulation patterns characteristic of $|B_{\rm{dip}}| = 0.33$ G (panel B) are already markedly different from the $|B_{\rm{dip}}| = 0.025$ G case (panel A), which clearly resembles the unmagnetized circulation. The flow is in essence entirely azimuthal once the field is increased to $|B_{\rm{dip}}| = 0.5$ G (panel C). This is in contrast to the non-uniformly dragged simulations of \citet{2013ApJ...764..103R}, who find the flow to become less zonally-dominated with enhanced field strength.

It is noteworthy that the flow speeds up once it takes on a zonal nature. This is almost certainly due to the fact that viscosity acting on vertical motion more strongly affects the divergent flow, and is therefore not a physically significant result. Increasing the field strength further diminished the flow velocities but did not alter the qualitative nature of the solution, although somewhat higher values of the Ekman number were required to ensure numerical stability. 

At the expense of a great inflation in the required computational time, we have extended the simulations presented in Figure (\ref{mhdsims}) to higher resolution. Namely, we prolonged the spherical harmonic decomposition up to degree $\ell_{\rm{max}} = 34$ and $m_{\rm{max}} = 29$ while eliminating hyperviscosity from our runs entirely. Aside from a mild (i.e. few percent) increase in the velocities, the results observed in these simulations were largely unchanged from the nominal simulations. This implies that the presented solutions do not depend sensitively on small-scale flows. In other words, the transition of the atmosphere to a state dominated by zonal jets is a result of interactions between global circulation and the large-scale magnetic field.

Provided the zonal nature of the flow observed in the magnetized simulation, we can expect that the induced field will be almost entirely toroidal and will approximately be described by equations (\ref{Bindtheta}) and (\ref{BindthetaA}). As shown in Figure (\ref{inducedfield}), this indeed appears to be the case. The numerically obtained azimuthal component of the field (panel B) is qualitatively similar to its analytically computed counterpart (panel A), although the field lines are concentrated towards the vicinity of the equator in the numerical solution (this is simply a consequence of the fact that the circulation is not exactly given by the expression (\ref{vzonal})). The magnitude of the induced field is also in good agreement with the analytical theory. For $\tilde{v}_0 = 440$ m/s, and $k_m/\mathcal{R}^3 = 0.5$ G, equation (\ref{Bindtheta}) yields $\rm{max}(B_{\rm{ind}}) = 0.64$ G, where as the numerically computed value is $\rm{max}(B_{\rm{ind}}) = 0.52$ G. 

Unlike the case considered in the previous section (where the Lorentz force is treated as a drag), within the framework of MHD simulations, the relationship between the peak wind speed and the temperature perturbation is not necessarily simple. Consequently, in order to preliminarily explore the sensitivity of our results on irradiation, we performed an additional suite of simulations where the applied heat flux was enhanced by a factor of three, compared to the simulations shown in Figure (\ref{mhdsims}). In these overdriven simulations, we found the peak wind speeds to be a factor of $\sim 2 - 2.5$ higher. However, the characteristic flow patterns closely resembled those, shown in Figure (\ref{mhdsims}) and specifically, the circulation with $|B_{\rm{dip}}| = 0.5$ G remained dominated by azimuthal jets. Consequently, we conclude that the transition of the circulation to a zonal regime with increasing magnetic field strength is robust within the context of our model.

It is interesting to note that the zonal nature of the circulation is ensured at a comparatively low magnetic field strength. If we adopt a scaling based on an Elsasser number of order unity \citep{2003E&PSL.208....1S}, typical hot Jupiter magnetic fields should exceed that of Jupiter by a factor of a few e.g. $|B_{\rm{dip}}| \sim 10$ G. Moreover, the arguably more physically sensible scaling based on the intrinsic energy flux \citep{2009Natur.457..167C} suggests that typical hot Jupiter fields may be still higher by another factor of $\sim 5$. Cumulatively, this places the critical magnetic field needed for the onset of zonal flows a factor of $\sim 10 - 100$ below the typical field strengths. As a result, it would be surprising if a more sophisticated treatment of the hydrodynamics and radiative transfer proved the critical field strength to be higher than the typical one. Nevertheless, such simulations should no doubt be performed. 

\section{Discussion}

In this paper, we have characterized the nature of atmospheric circulation on Hot Jupiters, in a regime where magnetic effects play an appreciable role. We began by performing baseline Boussinesq hydrodynamical simulations and augmenting them to crudely account for the Lorentz force by expressing it in the form of a Rayleigh drag. Using a simple analytical theory, we showed that within the framework of such a treatment, the interactions between the circulation and the background magnetic field lead to a well-formulated reduction in wind velocities. However, in agreement with published literature \citep{2010ApJ...719.1421P,2012ApJ...750...96R} and dragged simulations of our own, we noted that regardless of the background field strength, the functional form of the upper-atmospheric stream function remains characteristic of a flow pattern where wind originates at the substellar point and blows towards the anti-stellar points quasi-symmetrically over the terminator (see also \citet{2008ApJ...685.1324S}).

Although simplifying, the assumption that the Lorentz force (even approximately) opposes the flow everywhere, as done by a Rayleigh drag, appears inappropriate for dayside-to-nightside circulation. Consequently, relying on theoretical considerations based on a kinematic treatment of magnetic induction \citep{1978mfge.book.....M}, we showed that if the Lorentz force is not reduced to a form of a drag, dayside-to-nighside flows become unstable in presence of a spin pole aligned magnetic field. On the contrary, the interactions between zonal jets and the background magnetic field do not give rise to meridional or radial flows, thanks to an inherent symmetry. Instead, the jets are stably damped by the background field \citep{2008Icar..196..653L, 2012ApJ...745..138M}. However, the damping rate generally need not be latitudinally uniform. 

As demonstrated by magnetohydrodynamical simulations, this has profound implications for upper-atmospheric circulation. Specifically, the MHD calculations indicate that once the background magnetic field is stronger than a critical value, the upper atmospheric circulation transitions from a state dominated by dayside-to-nightside flows to an azimuthally symmetric pattern dominated by zonal jets. Qualitatively, this transition can be understood as a point where redistribution of heat from the dayside to the nightside by flow patterns that intersect the magnetic poles ceases to be energetically favorable against purely zonal circulation. For the standard case considered here (that is, $\tau_{\rm{N}}= 10^3$ sec), the critical field strength is approximately $B_{\rm{crit}} \simeq 0.5$ G, considerably less than the typically inferred field strengths of Hot Jupiters \citep{2009Natur.457..167C}. However, it should be understood that the critical field strength must unavoidably depend on various system parameters including the radiative timescale and the electrical conductivity. The variability due to the latter may be particularly important since thermal ionization is extremely sensitive to the atmospheric temperature (see Figure \ref{sigma}). The form of this dependence and the extent to which atmospheres within the current observational aggregate are magnetically dominated merits further investigation.

The fact that dayside-to-nightside flows tend to simplify to a zonal state in magnetized atmospheres has a number of important implications. As already discussed to some extent in section (3), axisymmetric flows give rise to exclusively toroidal fields \citep{1978mfge.book.....M}. This means that additional atmospheric dynamo generation, that would act to augment a deep seated field, cannot be supported by large-scale circulation. Moreover, because the induced field lacks a strong poloidal component, its observational characterization is at present hopeless. Consequently, observational inference of magnetohydrodynamic processes in exoplanetary atmospheres is likely to be limited to indirect methods.

This discussion overlooks the possibility of field generation by small-scale turbulence (i.e. the $\alpha-$effect) in the atmosphere. Indeed such a process may be at play if the turbulent magnetic Reynolds number $Re_m^t \equiv \nu/\eta \gtrsim 1 - 10$. For highly turbulent atmospheres, this criterion may indeed be satisfied. Our simulations aimed at determining the viability as well as characterization of field generation by small-scale turbulence in Hot Jupiter atmospheres are currently ongoing and will be reported in a follow-up study.

In this work, we briefly hinted at the fact that the damping of zonal jets by dipole magnetic fields is not only non-uniform latitudinally but also radially. The radial dependence of the Lorentz force found here is specific to the boundary conditions imposed on the induced current. However, if we do not choose to confine the current to a single scale-height but allow it to penetrate the convective interior of the planet (as for example envisioned within the context of the \citet{2010ApJ...714L.238B} Ohmic dissipation model), the induced toroidal field is free to occupy a much deeper portion of the planet. In such a case, the resulting Lorentz force may act to produce deep-seated azimuthal flows and give rise to large-scale differential rotation within the planet \citep{Peterchat}. However, the extent to which such differential rotation can persist is subject to a number of constraints, including the magnitude of interior Ohmic dissipation \citep{2008Icar..196..653L}.

In addition to the various simplifications inherent to our model that are already described throughout the paper, it is further noteworthy that we have restricted the morphology of the background magnetic field to a pole-aligned dipole for simplicity. Within the solar system, a dipolar, axisymmetric magnetic field created by an internal dynamo is possessed only by Saturn \citep{1980Sci...207..444A,2005Sci...307.1266D}. On the contrary, Jupiter and the Earth have dipole-dominated fields that are significantly tilted with respect to their spin-axes, while Neptune and Uranus possess rather unusual non-dipolar, non-axisymmetric fields \citep{2003E&PSL.208....1S}. As a result it is quite likely that even on a qualitative level, the discussion presented in this work is not comprehensive. That is, unlike axisymmetric jets found in this work, one could envision the generation of substantial stationary eddies, yielding longitudinally and latitudinally asymmetric jets in exoplanetary atmospheres, by complex background magnetic fields. 

Furthermore, orbital variations may also be of importance. Specifically, while the assumption of a circular orbit is secure for the majority of hot Jupiters, null eccentricities are certainly not universal to the observational sample (an extreme example is HD80606b \citep{2001A&A...375..205N} which has $e=0.93$). The time-variability of stellar irradiation associated with significant eccentricity produces rather complex circulation patterns even in hydrodynamic regime \citep{2008ApJ...674.1106L, 2013ApJ...767...76K}. However, recalling that electrical conductivity in hot planetary atmospheres arises primarily as a result of thermal ionization, the circulation patterns are likely to be even more complex than those typically envisioned, since magnetic effects in the atmosphere will also be time-dependent.

We would like to finish with a few words about observational implications of our results. At present, the resolution and signal to noise of the spectroscopic data aimed at characterizing the temperature structure and chemical composition of exoplanetary atmospheres \citep{2005ApJ...626..523C, 2008ApJ...673..526K, 2010Natur.463..637S} is such that even fits obtained with one-dimensional atmospheric models are susceptible to numerous degeneracies \citep{2009ApJ...707...24M}. Consequently, it is unlikely that the qualitative change in the flow structure observed in this work will strongly affect the already-limited interpretation of the information contained within this data, in the near future \citep{2012ApJ...749...93L}. 

On the other hand, theoretical interpretation of observed dayside-to-nightside temperature differences and the associated shifts in the location of the subsolar hot spot \citep{2007Natur.448..143K} rely heavily on a sensible understanding of atmospheric dynamics, which as we have seen requires a more or less self-consistent account of magnetic effects. To this end, the results obtained in this study are of great importance. Indeed, one can expect that the advective transport of heat changes character and weakens with increased electrical conductivity (by extension, the atmospheric temperature) and magnetic field due to the mechanism described above (see also \citet{2008Icar..196..653L, 2010ApJ...719.1421P, 2011ApJ...738....1B, 2012ApJ...754L...9M, 2012ApJ...750...96R}). Thus, a thorough comparison between a substantial sample of model results and data may eventually shed light on the typical atmospheric conductivity structure and field strengths of Hot Jupiters. Such activity would no-doubt further benefit from direct measurements of high-altitude atmospheric wind velocities obtained via ground-based spectroscopy \citep{2010Natur.465.1049S}. That said, in order for an endeavor of this sort to be meaningful, substantial improvements in theoretical modeling aimed at meliorating the shortcomings outlined above are required, along with a wealth of additional observational data.

In conclusion, the above discussion clearly indicates that the degree of complexity of the physical regime in which hot exoplanetary atmospheres reside is indeed very extensive. There is no doubt that much additional work remains. In this work, we have taken an ample step towards a self-consistent characterization of magnetically controlled circulation on Hot Jupiters. As such, this study should serve as a stepping stone for future developments.

\acknowledgments 
\textbf{Acknowledgments}. We thank Adam Showman, Tami Rogers, Kristen Menou, Peter Goldreich and Greg Laughlin for useful conversations, as well as the anonymous referee for a thorough and insightful report. K.B. acknowledges the generous support from the ITC Prize Postdoctoral Fellowship at the Institute for Theory and Computation, Harvard-Smithsonian Center for Astrophysics. S.S. acknowledges funding by the Natural Sciences and Engineering Research Council of Canada (NSERC) and the Alfred P. Sloan Foundation.


\begin{thebibliography} 

\bibitem[Acuna 
\& Ness(1980)]{1980Sci...207..444A} Acuna, M.~H., \& Ness, N.~F.\ 1980, Science, 207, 444 

\bibitem[Batalha et al.(2012)]{2012arXiv1202.5852B} Batalha, N.~M., Rowe, 
J.~F., Bryson, S.~T., et al.\ 2012, arXiv:1202.5852 

\bibitem[Batygin 
\& Stevenson(2010)]{2010ApJ...714L.238B} Batygin, K., \& Stevenson, D.~J.\ 2010, \apjl, 714, L238 

\bibitem[Batygin et al.(2011)]{2011ApJ...738....1B} Batygin, K., Stevenson, 
D.~J., \& Bodenheimer, P.~H.\ 2011, \apj, 738, 1 

\bibitem[Burrows et al.(2007)]{2007ApJ...661..502B} Burrows, A., Hubeny, 
I., Budaj, J., \& Hubbard, W.~B.\ 2007, \apj, 661, 502 

\bibitem[Burrows et al.(2007)]{2007ApJ...668L.171B} Burrows, A., Hubeny, 
I., Budaj, J., Knutson, H.~A., \& Charbonneau, D.\ 2007, \apjl, 668, L171 

\bibitem[Cebron et 
al.(2012)]{2012A&A...539A..78C} Cebron, D., Le Bars, M., Moutou, C., \& Le Gal, P.\ 2012, \aap, 539, A78 

\bibitem[Charbonneau et al.(2000)]{2000ApJ...529L..45C} Charbonneau, D., 
Brown, T.~M., Latham, D.~W., \& Mayor, M.\ 2000, \apjl, 529, L45 

\bibitem[Charbonneau et al.(2005)]{2005ApJ...626..523C} Charbonneau, D., 
Allen, L.~E., Megeath, S.~T., et al.\ 2005, \apj, 626, 523 

\bibitem[Cho et al.(2003)]{2003ApJ...587L.117C} Cho, J.~Y.-K., Menou, K., 
Hansen, B.~M.~S., \& Seager, S.\ 2003, \apjl, 587, L117 

\bibitem[Cho et al.(2008)]{2008ApJ...675..817C} Cho, J.~Y.-K., Menou, K., 
Hansen, B.~M.~S., \& Seager, S.\ 2008, \apj, 675, 817 

\bibitem[Christensen 
\& Wicht(2008)]{2008Icar..196...16C} Christensen, U.~R., \& Wicht, J.\ 2008, Icarus, 196, 16 

\bibitem[Christensen et al.(2009)]{2009Natur.457..167C} Christensen, U.~R., 
Holzwarth, V., \& Reiners, A.\ 2009, \nat, 457, 167 

\bibitem[Clayton(1968)]{1968psen.book.....C} Clayton, D.~D.\ 1968, New 
York: McGraw-Hill, 1968

\bibitem[Cooper 
\& Showman(2005)]{2005ApJ...629L..45C} Cooper, C.~S., \& Showman, A.~P.\ 2005, \apjl, 629, L45 

\bibitem[Dobbs-Dixon 
\& Lin(2008)]{2008ApJ...673..513D} Dobbs-Dixon, I., \& Lin, D.~N.~C.\ 2008, \apj, 673, 513

\bibitem[Dobbs-Dixon 
\& Agol(2012)]{2012arXiv1211.1709D} Dobbs-Dixon, I., \& Agol, E.\ 2012, arXiv:1211.1709

\bibitem[Dougherty et al.(2005)]{2005Sci...307.1266D} Dougherty, M.~K., 
Achilleos, N., Andre, N., et al.\ 2005, Science, 307, 1266 

\bibitem[Dumberry 
\& Bloxham(2002)]{2002GeoJI.151..377D} Dumberry, M., \& Bloxham, J.\ 2002, Geophysical Journal International, 151, 377 

\bibitem[Goldreich(private communication)]{Peterchat} Goldreich, P., \ 2011, private communication 

\bibitem[Hallinan et al.(2013)]{2013ApJ...762...34H} Hallinan, G., Sirothia, S.~K., Antonova, A., et al.\ 2013, \apj, 762, 34 

\bibitem[Heng et al.(2011)]{2011MNRAS.tmp..370H} Heng, K., Menou, K., 
\& Phillipps, P.~J.\ 2011, \mnras, 370 

\bibitem[Heng et al.(2011)]{2011MNRAS.418.2669H} Heng, K., Frierson, 
D.~M.~W., \& Phillipps, P.~J.\ 2011, \mnras, 418, 2669 

\bibitem[Heng(2012)]{2012ApJ...748L..17H} Heng, K.\ 2012, \apjl, 748, L17 

\bibitem[Held \& Hou(1980)]{1980JAtS...37..515H} Held, I.~M., \& Hou, A.~Y.\ 1980, Journal of Atmospheric Sciences, 37, 515 

\bibitem[Holton(1992)]{1992aitd.book.....H} Holton, J.~R.\ 1992, 
International geophysics series, San Diego, New York: Academic Press, 
|c1992, 3rd ed.

\bibitem[Guillot 
\& Showman(2002)]{2002A&A...385..156G} Guillot, T., \& Showman, A.~P.\ 2002, \aap, 385, 156 

\bibitem[Henry et al.(2000)]{2000ApJ...529L..41H} Henry, G.~W., Marcy, 
G.~W., Butler, R.~P., \& Vogt, S.~S.\ 2000, \apjl, 529, L41 

\bibitem[Hut(1981)]{1981A&A....99..126H} Hut, P.\ 1981, \aap, 99, 126 

\bibitem[Kataria et al.(2013)]{2013ApJ...767...76K} Kataria, T., Showman, 
A.~P., Lewis, N.~K., et al.\ 2013, \apj, 767, 76 

\bibitem[Knutson(2007)]{2007Natur.448..143K} Knutson, H.~A.\ 2007, \nat, 
448, 143 

\bibitem[Knutson et al.(2008)]{2008ApJ...673..526K} Knutson, H.~A., 
Charbonneau, D., Allen, L.~E., Burrows, A., 
\& Megeath, S.~T.\ 2008, \apj, 673, 526

\bibitem[Knutson et al.(2009)]{2009ApJ...691..866K} Knutson, H.~A., 
Charbonneau, D., Burrows, A., O'Donovan, F.~T., 
\& Mandushev, G.\ 2009, \apj, 691, 866 

\bibitem[Kuang 
\& Bloxham(1999)]{1999JCoPh.153...51K} Kuang, W., \& Bloxham, J.\ 1999, Journal of Computational Physics, 153, 51

\bibitem[Jackson(1998)]{1998clel.book.....J} Jackson, J.~D.\ 1998, 
Classical Electrodynamics, 3rd Edition, by John David Jackson, 
pp.~832.~ISBN 0-471-30932-X.~Wiley-VCH , July 1998.,

\bibitem[Landau 
\& Lifshitz(1959)]{1959flme.book.....L} Landau, L.~D., \& Lifshitz, E.~M.\ 1959, Course of theoretical physics, Oxford: Pergamon Press, 1959, 

\bibitem[Langton 
\& Laughlin(2007)]{2007ApJ...657L.113L} Langton, J., \& Laughlin, G.\ 2007, \apjl, 657, L113 

\bibitem[Langton 
\& Laughlin(2008)]{2008ApJ...674.1106L} Langton, J., \& Laughlin, G.\ 2008, \apj, 674, 1106 

\bibitem[Lodders(1999)]{1999ApJ...519..793L} Lodders, K.\ 1999, \apj, 519, 
793 

\bibitem[Line et al.(2012)]{2012ApJ...749...93L} Line, M.~R., Zhang, X., 
Vasisht, G., et al.\ 2012, \apj, 749, 93 

\bibitem[Liu et al.(2008)]{2008Icar..196..653L} Liu, J., Goldreich, P.~M., \& Stevenson, D.~J.\ 2008, icarus, 196, 653 

\bibitem[Liu 
\& Showman(2012)]{2012arXiv1208.0126L} Liu, B., \& Showman, A.~P.\ 2012, arXiv:1208.0126 

\bibitem[Madhusudhan 
\& Seager(2009)]{2009ApJ...707...24M} Madhusudhan, N., \& Seager, S.\ 2009, \apj, 707, 24 

\bibitem[Marcy 
\& Butler(1996)]{1996ApJ...464L.147M} Marcy, G.~W., \& Butler, R.~P.\ 1996, \apjl, 464, L147 

\bibitem[Mayor 
\& Queloz(1995)]{1995Natur.378..355M} Mayor, M., \& Queloz, D.\ 1995, \nat, 378, 355 

\bibitem[Menou 
\& Rauscher(2009)]{2009ApJ...700..887M} Menou, K., \& Rauscher, E.\ 2009, \apj, 700, 887 

\bibitem[Menou 
\& Rauscher(2010)]{2010ApJ...713.1174M} Menou, K., \& Rauscher, E.\ 2010, \apj, 713, 1174 

\bibitem[Menou(2012)]{2012ApJ...754L...9M} Menou, K.\ 2012, \apjl, 754, L9 

\bibitem[Menou(2012)]{2012ApJ...745..138M} Menou, K.\ 2012, \apj, 745, 138 

\bibitem[Moffatt(1978)]{1978mfge.book.....M} Moffatt, H.~K.\ 1978, 
Cambridge, England, Cambridge University Press, 1978.~353 p.

\bibitem[Naef et 
al.(2001)]{2001A&A...375..205N} Naef, D., Mayor, M., Pepe, F., et al.\ 2001, \aap, 375, 205 

\bibitem[Perna et al.(2010)]{2010ApJ...719.1421P} Perna, R., Menou, K., 
\& Rauscher, E.\ 2010, \apj, 719, 1421 

\bibitem[Peixoto 
\& Oort(1992)]{1992phcl.book.....P} Peixoto, J.~P., \& Oort, A.~H.\ 1992, New York: American Institute of Physics (AIP), 1992

\bibitem[Rauscher 
\& Menou(2012)]{2012ApJ...750...96R} Rauscher, E., \& Menou, K.\ 2012, \apj, 750, 96 

\bibitem[Rauscher 
\& Menou(2013)]{2013ApJ...764..103R} Rauscher, E., \& Menou, K.\ 2013, \apj, 764, 103

\bibitem[Rodgers(2000)]{2000SAOPP...2.....R} Rodgers, C.~D.\ 2000, Inverse 
Methods for Atmospheric Sounding - Theory and Practice.~Series: Series on 
Atmospheric Oceanic and Planetary Physics, ISBN: <ISBN>9789812813718</ISBN>.~World 
Scientific Publishing Co.~Pte.~Ltd., Edited by Clive D.~Rodgers, vol.~2, 2, 

\bibitem[Sarson et al.(1997)]{1997Sci...276.1106S} Sarson, G.~R., Jones, 
C.~A., Zhang, K., \& Schubert, G.\ 1997, Science, 276, 1106 

\bibitem[Schneider 
\& Lindzen(1977)]{1977JAtS...34..263S} Schneider, E.~K., \& Lindzen, R.~S.\ 1977, Journal of Atmospheric Sciences, 34, 263 

\bibitem[Showman 
\& Guillot(2002)]{2002A&A...385..166S} Showman, A.~P., \& Guillot, T.\ 2002, \aap, 385, 166 

\bibitem[Showman et al.(2008)]{2008ApJ...685.1324S} Showman, A.~P., Cooper, 
C.~S., Fortney, J.~J., \& Marley, M.~S.\ 2008, \apj, 685, 1324 

\bibitem[Showman et al.(2009)]{2009ApJ...699..564S} Showman, A.~P., Fortney, J.~J., Lian, Y., et al.\ 2009, \apj, 699, 564 

\bibitem[Showman \& Polvani(2011)]{2011ApJ...738...71S} Showman, A.~P., \& Polvani, L.~M.\ 2011, \apj, 738, 71 

\bibitem[Showman et al.(2011)]{2011exop.book..471S} Showman, A.~P., Cho, 
J.~Y.-K., \& Menou, K.\ 2011, Exoplanets, edited by S.~Seager.~ Tucson, AZ: University of Arizona Press, 2011, 526 pp.~ ISBN 978-0-8165-2945-2., p.471-516, 471 

\bibitem[Snellen et al.(2010)]{2010Natur.465.1049S} Snellen, I.~A.~G., de 
Kok, R.~J., de Mooij, E.~J.~W., \& Albrecht, S.\ 2010, \nat, 465, 1049 

\bibitem[Spiegel et al.(2009)]{2009ApJ...699.1487S} Spiegel, D.~S., 
Silverio, K., \& Burrows, A.\ 2009, \apj, 699, 1487 

\bibitem[Stanley 
\& Bloxham(2004)]{2004Natur.428..151S} Stanley, S., \& Bloxham, J.\ 2004, \nat, 428, 151

\bibitem[Stanley 
\& Bloxham(2006)]{2006Icar..184..556S} Stanley, S., \& Bloxham, J.\ 2006, Icarus, 184, 556 

\bibitem[Stanley et 
al.(2005)]{2005E&PSL.234...27S} Stanley, S., Bloxham, J., Hutchison, W.~E., \& Zuber, M.~T.\ 2005, Earth and Planetary Science Letters, 234, 27 

\bibitem[Stanley 
\& Mohammadi(2008)]{2008PEPI..168..179S} Stanley, S., \& Mohammadi, A.\ 2008, Physics of the Earth and Planetary Interiors, 168, 179 

\bibitem[Stanley et al.(2008)]{2008Sci...321.1822S} Stanley, S., 
Elkins-Tanton, L., Zuber, M.~T., 
\& Parmentier, E.~M.\ 2008, Science, 321, 1822 

\bibitem[Stanley(2010)]{2010GeoRL..37.5201S} Stanley, S.\ 2010, \grl, 37, 
5201

\bibitem[Stevenson(2003)]{2003E&PSL.208....1S} Stevenson, D.~J.\ 2003, Earth and Planetary Science Letters, 208, 1 

\bibitem[Swain et al.(2010)]{2010Natur.463..637S} Swain, M.~R., Deroo, P., 
Griffith, C.~A., et al.\ 2010, \nat, 463, 637 

\bibitem[Iro et 
al.(2005)]{2005A&A...436..719I} Iro, N., B{\'e}zard, B., \& Guillot, T.\ 2005, \aap, 436, 719 

\bibitem[Youdin 
\& Mitchell(2010)]{2010ApJ...721.1113Y} Youdin, A.~N., \& Mitchell, J.~L.\ 2010, \apj, 721, 1113

\bibitem[Zuber et al.(2007)]{2007SSRv..131..105Z} Zuber, M.~T., Aharonson, 
O., Aurnou, J.~M., et al.\ 2007, \ssr, 131, 105 

\end{thebibliography}
\end{document}